\documentclass[aps,pre,reprint,showpacs,superscriptaddress,longbibliography]{revtex4-2}
\usepackage{mathtools,amssymb,graphicx,units}
\usepackage[usenames,dvipsnames]{color}
\usepackage[plainpages=false,pdfpagelabels,colorlinks=true,linkcolor=red,urlcolor=blue,citecolor=blue,pdftitle={Title},pdfauthor={},pdfdisplaydoctitle=true,pdfduplex=DuplexFlipLongEdge]{hyperref}
\usepackage{subfigure}
\usepackage{grffile}
\usepackage{bm}
\usepackage{mhchem}
\usepackage{bibentry}

\newcommand{\abs}[1]{\ensuremath{\left| #1 \right|}}

\newcommand{\vev}[1]{\left\langle #1\right\rangle}
\newcommand{\1}{\mbox{\bf 1}}
\newcommand\redsout{\bgroup\markoverwith{\textcolor{red}{\rule[0.5ex]{2pt}{0.4pt}}}\ULon}

\begin{document}

%\preprint{}

\title{Characterizing spin ordering via maximal row correlation in classical spin models}

\author{Yong-Yi Tang}
\affiliation{Lanzhou Center for Theoretical Physics, Key Laboratory of Quantum Theory and Applications of MoE, and Key Laboratory of Theoretical Physics of Gansu Province, Lanzhou University, Lanzhou 730000, China}
\author{Yin Zhong}
\affiliation{Lanzhou Center for Theoretical Physics, Key Laboratory of Quantum Theory and Applications of MoE, and Key Laboratory of Theoretical Physics of Gansu Province, Lanzhou University, Lanzhou 730000, China}
\author{Hantao Lu}
\email{luht@lzu.edu.cn}
\affiliation{Lanzhou Center for Theoretical Physics, Key Laboratory of Quantum Theory and Applications of MoE, and Key Laboratory of Theoretical Physics of Gansu Province, Lanzhou University, Lanzhou 730000, China}

\date{\today}

\begin{abstract}
An order parameter, termed the maximal row correlation, is proposed for classical spin systems. Monte Carlo simulations on various Potts models suggest that this order parameter is applicable to a broad range of spin systems, including those defined on irregular lattices, systems with frustration, and systems exhibiting partial orders, provided some degree of spin ordering is present. This approach offers a unified framework for investigating phase transitions in such complex systems. The associated critical exponents are estimated via finite-size scaling analysis and show good agreement with established values.
\end{abstract}

%\keywords{}

\maketitle

\section{Introduction}\label{sec_intro}

In Landau's theory of phase transitions~\cite{Landau_1980}, the concept of spontaneous symmetry breaking plays a central role and is characterized by the emergence or disappearance of certain orders. In many cases, the corresponding order parameter can be readily identified, provided that the broken symmetry is known in advance. However, there are situations where the appropriate order parameters are elusive, such as in the case of partial ordering in Potts models with antiferromagnetic (AFM) interactions on irregular lattices~\cite{ChenQN_2011}. 

The Potts model, introduced by R. B. Potts in his 1951 PhD thesis~\cite{Potts_1951} and subsequent 1952 paper~\cite{Potts_1952}, is a fundamental model in statistical physics. It generalizes the Ising model by allowing each lattice site to occupy one of $q$ discrete states. Over the past half-century, it has been the subject of extensive study. For comprehensive reviews, see, e.g., Refs.~\cite{WuFY_1982} and ~\cite{Baxter_1982a}. The phase transition of the Potts model with ferromagnetic (FM) interaction is relatively straightforward, and its properties are well understood. In contrast, the behavior of the AFM Potts model is significantly more complex, with phase transition characteristics that depend on both the lattice structure and the number of states $q$. Some Potts models exhibit macroscopic ground-state degeneracy and zero-temperature entropy, potentially giving rise to phases where certain sites are ordered while others remain disordered. This phenomenon of partial order has been observed in both frustrated~\cite{Lipowski_1995,Foster_2004,Igarashi_2010,Qin_2013} and non-frustrated models~\cite{Parsonage_1978,ChenQN_2011}, highlighting the difficulty of defining a clear order parameter for the AFM Potts model. 

A common approach for studying partial orders is to compute the partial or staggered magnetization by adding or subtracting the magnetization of sublattices, which often requires model-specific analysis~\cite{Kotecky_2008,Tseng_2024}. For certain models, however, identifying the order from low-temperature configurations can be challenging, as seen in the case of the $q=2$ AFM Ising model on the Union-Jack lattice (see also Fig.~\ref{fig_union_jack} in the main text).

Recently, in contrast to conventional methods of identifying phase transitions and critical points through the study of thermodynamic quantities and order parameters, substantial progress has been made in exploring alternative approaches. These include the ``pattern picture"~\cite{Yang_2022}, the perspective of information theory~\cite{Prokopenko_2011}, analyzing configuration correlations during the Monte Carlo (MC) process~\cite{ChenXS_2019,Cheng_2023,*Cheng_2024}, and machine learning-based analyses~\cite{Santos_2021a,Miyajima_2023}, among others. Inspired by these advancements, the present study aims to propose a unified and physically sound order parameter within a general framework to describe the phase transitions in various Potts models, particularly those with AFM interactions and no apparent definition of long-range orders. 

In the paper, we introduce an order parameter termed the {\it maximal row correlation}, denoted by $O$, which can be evaluated from the row correlation matrix. We perform classical MC simulations on various Potts models defined on both regular and irregular two-dimensional (2D) lattices, computing the temperature dependence of $O$ and its fluctuations, $\delta O$. For systems with FM interactions, we demonstrate that the behavior of $O$ and $\delta O$ closely parallels that of conventional magnetization and susceptibility, respectively. Remarkably, for models with either unfrustrated or frustrated AFM interactions, these quantities can serve as effective indicators of phase transition without the need for modification. In unfrustrated cases, we estimate the critical temperature and the corresponding critical exponents via finite-size scaling analysis of $O$, obtaining results that are in good agreement with established values.

The remainder of the paper is organized as follows. In Sec.~\ref{sec_methods}, we define the maximal row correlation $O$, discuss its physical significance, and describe the MC simulation procedures in which $O$ serves as the order parameter. In Sec.~\ref{sec_results}, we benchmark the effectiveness of $O$ in phase transitions of spin models by applying it to the $q=2$ and $q=3$ FM Potts models on the square lattice. We then demonstrate its advantages in AFM systems, including the $q=3$ on the diced lattice and the $q=2$, $q=3$ models on the Union-Jack lattice. A summary and discussion are presented in Sec.~\ref{sec_summary}. Additional details of the order parameter, including its behavior in high- and low-temperature limits, and its potential extensions to higher-dimensional and continuous-spin models, are provided in the Appendix.

\section{Methods}\label{sec_methods}
\subsection{Maximal row correlation}\label{subsec_methods_O}

\begin{figure*}[htbp]
\centering
\includegraphics[width=\textwidth]{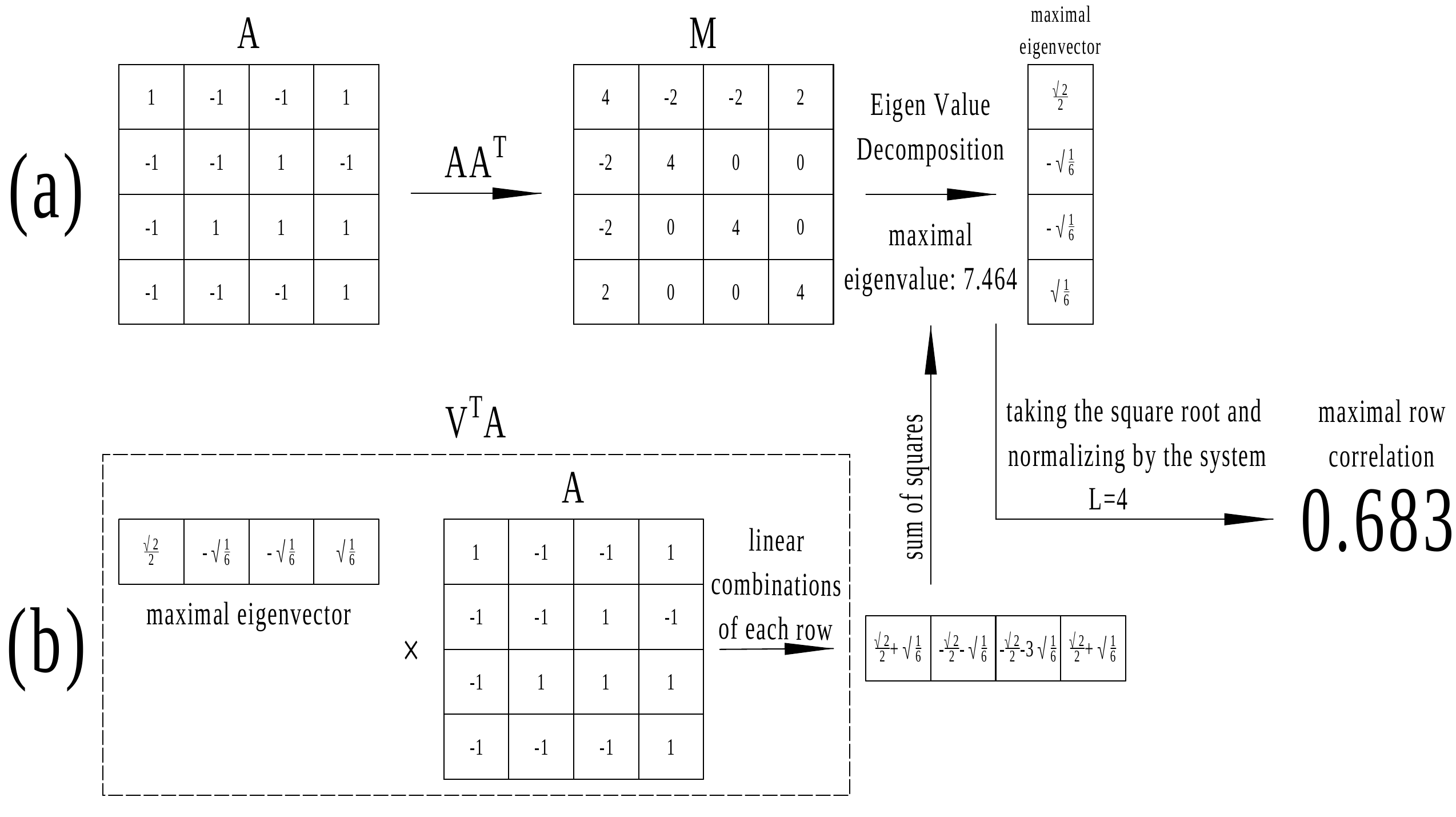}
\caption{(a) An example of the configuration matrix $A$ corresponding to a spin configuration of the Ising model on the $4\times 4$ square lattice, where the entries $+1/-1$ represent the spin orientations (up and down, respectively) at the lattice sites. The maximal row correlation is calculated from the largest eigenvalue of the row correlation matrix $M=A\,A^{\mathrm{T}}$. (b) The purpose of the process is to linearly combine the rows of the matrix $A$ to extract the maximal degree of order present in the spin configuration.}
\label{fig_process}
\end{figure*}

In this subsection, we give a detailed account of the order parameter of maximal row correlation proposed for the study. To illustrate the concept, we first apply it to the simple Ising model on a square lattice. The extension to more general systems is straightforward and will be addressed subsequently.

As illustrated in Fig.~\ref{fig_process}, any given spin configuration on the square lattice can be encoded in a matrix $A$ of the same dimensions, where each matrix elements records the spin orientation at the corresponding lattice site. We refer to this as {\em the configuration matrix}. In principle, any 2D lattice can be rearranged into a rectangular array, allowing the application of the configuration matrix representation. For higher-dimensional lattices, the corresponding configuration matrix can be constructed by reshaping higher-order tensors, for example by combining multiple indices into a single one. As a illustration, we refer the reader to Appendix~\ref{app_highD}, where the 3D and 4D Potts models are discussed.

A positive semi-definite square matrix $M$ can be constructed from $A$ as $M=A\,A^{\mathrm{T}}$, with its dimension determined by the number of rows of $A$. The physical meaning of the matrix $M$ is clear: each entry $(M)_{ij}$ represents the inner product between the $i$-th and $j$-th rows of $A$, thereby quantifying their correlation. Therefore we call $M$ {\em the row correlation matrix}. The maximal row correlation, denoted as ${O}$ in the paper, is simply defined to be the square root of $\lambda_{\mathrm{M}}$ that is the largest eigenvalue of $M$, divided by the matrix size, i.e., $O=\sqrt{\lambda_{\mathrm{M}}}/L$ ($L$ refers to the number of rows of $A$), as illustrated in Fig.~\ref{fig_process}(a). Consequently, $\delta O$, which measures the fluctuation of $O$, is written as $\delta O=\left(\vev{O^2}-\vev{O}^2\right)\times L$. The reason for these rescalings is explained in Appendix~\ref{app_A}.

The geometric significance of $\lambda_{\mathrm{M}}$, the largest eigenvalue of the row correlation matrix $M$, is as follows. Let $\mathbf{v}$ be an eigenvector of $M$ with eigenvalue $\lambda$, such that $M\mathbf{v}=\lambda\mathbf{v}$. Then, $\mathbf{v}^{\mathrm{T}}A\,A^{\mathrm{T}}\mathbf{v}=\lambda\mathbf{v}^{\mathrm{T}}\mathbf{v}$. This relation implies that $\lambda_{\mathrm{M}}$ gives the maximal squared length of any linear combination of the row vectors of $A$ with normalized coefficients, i.e., under the constraint $\mathbf{v}^{\mathrm{T}}\mathbf{v}=1$. The maximum is attained when $\mathbf{v}$ is the eigenvector of $M$ corresponding to $\lambda_{\mathrm{M}}$. This interpretation is illustrated in Fig.~\ref{fig_process}(b). That is why we refer to the order parameter $O$ as the {\em maximal row correlation}. It is worth noting that the row correlation matrix $M$, derived from an individual spin configuration, is distinct from the standard covariance matrix employed in principal component analysis (PCA)~\cite{Jolliffe_2002}, which is typically defined over an ensemble of configurations~\cite{WangLei_2016,ChenXS_2019}.

% The matrix $M$, defined as the product of the configuration matrix $A$ with its transpose $A^{\mathrm{T}}$, is analogous to the reduced density matrix in the density-matrix renormalization group (DMRG)~\cite{Schollwock_2005}, apart from the normalization condition. In this analogy, the configuration matrix $A$ corresponds to the superblock wavefunction in the DMRG framework, with its row and column indices representing the basis states of the system and environment, respectively. In this sense, the eigenvalue distribution of $M$ may be interpreted as encoding the entanglement characteristics of the spin configurations between the row and column directions. As for the largest eigenvalue of $M$, a geometric understanding is following: let $\mathbf{v}$ be an eigenvector of $M$ with eigenvalue $\lambda$, i.e., $M\mathbf{v}=\lambda\mathbf{v}$, then we have $\mathbf{v}^{\mathrm{T}}A\,A^{\mathrm{T}}\mathbf{v}=\lambda\mathbf{v}^{\mathrm{T}}\mathbf{v}$. This indicates that the largest eigenvalue of $M$ corresponds to the maximal squared length among all vectors obtained by linear combinations of the row vectors of $A$, where the coefficients are normalized. It occurs when the coefficient vector $\mathbf{v}$ coincides with the eigenvector of $M$ associated with the largest eigenvalue $\lambda_{\mathrm{M}}$. This interpretation is illustrated in Fig.~\ref{fig_process}(b). That is why we refer to the order parameter $O$ as the {\em maximal row correlation}.

Before concluding this subsection, we would like to discuss the generalization of the newly defined order parameter $O$ from the Ising model to the Potts model with $q$ states, where each entry in the configuration matrix $A$ can take $q$ distinct values, namely from $\{0,1,\ldots,q-1\}$ (see also Eq.~(\ref{eq_H_Potts})). In order to calculate the matrix $M$, the inner produce of two row vectors in $A$ can be defined as
\begin{equation}
\mathbf{v}\cdot{\mathbf{w}}=\sum_{i}v_i w_i,
\label{eq_innerproduct}
\end{equation}
with 
\begin{equation}
v_i w_i=\left\{\begin{array}{ll} 1, & v_i=w_i, \\ -\frac{1}{q-1}, & v_i\neq w_i. \end{array}\right.
\label{eq_spinproduct}
\end{equation}
By construction, the definition ensures that the average spin correlation between any two sites with random (i.e., uncorrelated) states is zero. When $q=2$, we can recover the familiar Ising case. Once $M$ is available, the order parameter $O$ can be calculated accordingly. As for the classical continuous spin model addressed in Appendix~\ref{app_continuous}, the component product between the spin configurations at two sites is defined as the inner (dot) product of their respective spin vectors. 

% For the three-state ferromagnetic Potts model on a square lattice, unlike the Ising model where spins can only take two values, \(\pm 1\), we treat the spins as vectors \((a, b)\), where \(a\) and \(b\) represent the components of the spin along the \(x\)- and \(y\)-axes, respectively. In this case, the original dot product between corresponding lattice sites in the Ising model is replaced by the inner product of vectors.

\subsection{Monte Carlo simulation} \label{subsec_methods_MC}

In the numerical simulations, for the Potts model with AFM interactions, the standard classical MC method with the Metropolis algorithm~\cite{Metropolis_1953} is employed. For each temperature, $200\times (q-1) \times L^2$ ($L$ is the system size) MC steps are performed for equilibration. In each MC step, a single-site spin-flip attempt is made. Each system size is simulated with $80$ independent annealing processes. In the afterward measurement phase, one sample is taken every $10\times L^2$ steps, yielding $1{,}000$ samples per process. This results in $80 \times 1{,}000$ configurations per temperature, from which the order parameter $O$ is computed for each configuration. These results are then evenly divided into $80$ bins for statistical analysis. In the vicinity of the critical regime, MC simulations are conducted with finer sampling to improve numerical precision. 

% These settings are sufficient to eliminate autocorrelation between consecutive MC steps and ensure ergodic sampling.

% (For high-dimensional models, e.g., in 3D and 4D cases, the configurations are instead divided into $100$ bins for analysis.)

On the other hand, for the Potts models with FM interactions, the Wolff update~\cite{Wolff_1989} is implemented near the critical point to improve equilibration efficiency, allowing equilibration with only $L^2$ cluster updates per temperature. In the measurement phase, one sample is taken every $10$ cluster updates. As in the AFM case, $80$ independent annealing processes are performed at each temperature, with $1{,}000$ samples collected per process.

In the presence of a phase transitions, analogous to the case of magnetization, the critical components---also denoted here as $\beta$ and $\nu$---along with the critical temperatures $T_{\mathrm{c}}$, are determined via a data collapse analysis of the maximal row correlation $O$. The procedure of data collapse is as follows. The data points for various lattice sizes near the critical regime are rescaled into a universal curve by choosing appropriate values of $T_{\mathrm{c}}$, $\beta$ and $\nu$ (for example, see Fig.~\ref{fig_square_fm_Q2}(c) for the case of the $q=2$ FM Potts model on the square lattice). The curve is fitted by a eleventh-order polynomial. We employ the binning method to estimate standard errors, and perform a chi-squared test~\cite{Landau_2024} based on these errors to assess the quality of the data collapse and to optimize the parameters $T_{\mathrm{c}}$, $\beta$, $\nu$. Note that the chi-squared value is defined as 
\begin{equation}
\chi^2 = \sum_{i=1}^{N} \frac{(y_i - f(x_i))^2}{\sigma_i^2},
\label{eq_chisquare}
\end{equation}
where $y_i$ denotes the scaled data points, $f(x_i)$ is the fitted curve, $\sigma_i$ is the standard error, and $N$ is the total number of data points. The uncertainties are estimated as the largest deviations of the parameters within the 68\% confidence interval, with additional contributions from the grid resolutions of $T_{\text{c}}$, $\beta$, and $\nu$ added in quadrature. 

To better access the quality of data fit relative to statistical uncertainty, in the following data collapse analysis we calculate the reduced chi-squared value as
\begin{equation}
\chi^2_{\text{red}} = \frac{\chi^2}{\text{dof}}.
\label{eq_chisquare_red}
\end{equation}
Here the degrees of freedom are given by $\text{dof}=N-p$, where $N$ is the number of data points and $p$ the number of fitting parameters. A value of $\chi^2_{\text{red}}$ close to unity typically indicates a good-quality fit. In our analysis, we use five system sizes for a given model, with $30$ data points per size near the critical temperature, yielding a total of $N=150$ data points. With the fitting curve described by an eleventh-order polynomial ($p=12$), this results in $\text{dof}=150-12=138$.

\section{Results}\label{sec_results}

In this section, we present the MC results on the maximal row correlation $O$ for the Potts model. The Hamiltonian for the Potts model in the absence of a magnetic field with isotropic interaction reads
\begin{equation}
H = -J \sum_{\langle i,j \rangle} \delta_{\sigma_i, \sigma_j}, \quad \sigma_i = 0, 1, \dots, q-1.
\label{eq_H_Potts}
\end{equation}
Here $q$ is the number of the available states for each site. $\sigma_i$ denotes the state at the $i$-th site, commonly referred to as the spin. $\delta_{\sigma_i, \sigma_j}$ is the Kronecker delta function, that takes the value of $1$ if $\sigma_i = \sigma_j$, and $0$ otherwise. $J$ is the coupling constant between spins. When $J>0$, the coupling is FM, whereas $J<0$ corresponds to AFM coupling. In this study, for simplicity, only the couplings between the nearest neighbors are considered. When $q=2$, the Potts model simplifies to the Ising model. In the following discussion, the magnitude of the coupling constant $J$ is always set to unity.

\subsection{$q=2$ and $q=3$ FM Potts model on the square lattice}\label{subsec_FM}

\begin{figure*}
\centering
\begin{minipage}{0.32\textwidth}
\centering
\includegraphics[width=\textwidth]{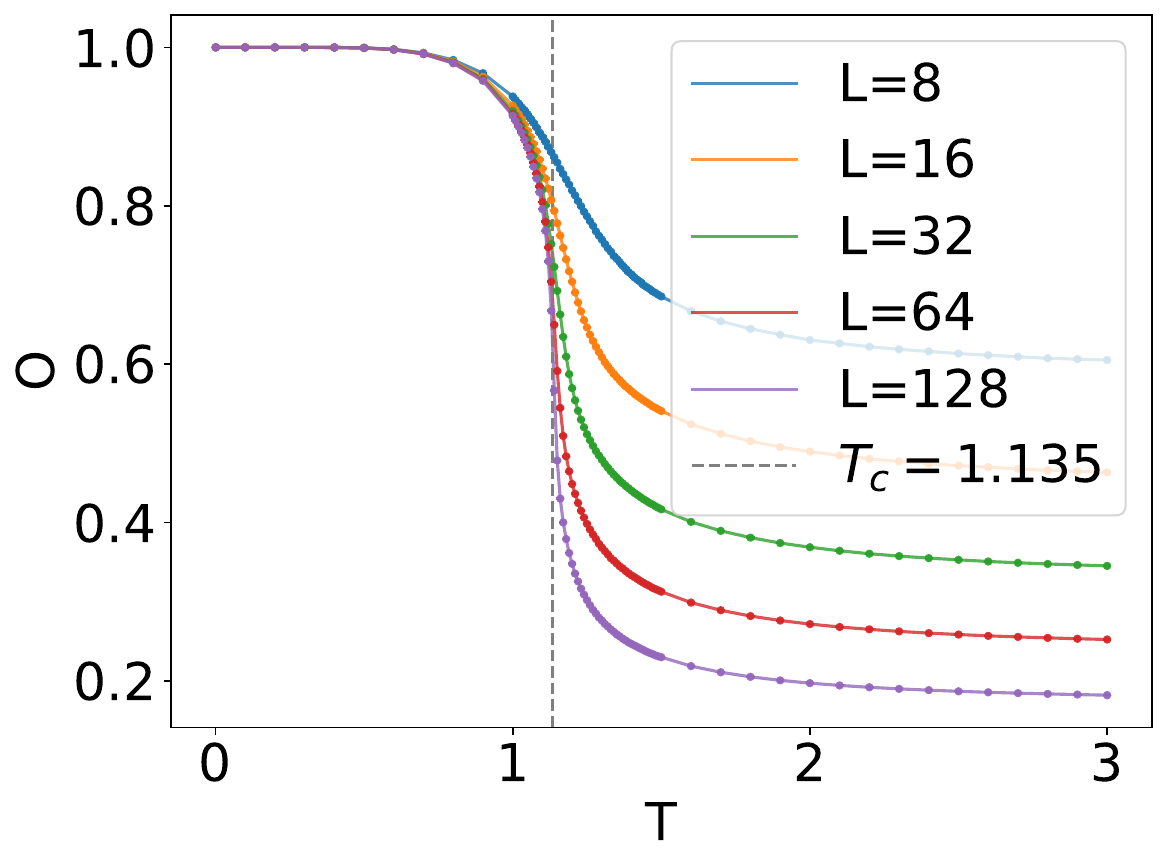}
\par
(a) 
\end{minipage}
\begin{minipage}{0.32\textwidth}
\centering
\includegraphics[width=\textwidth]{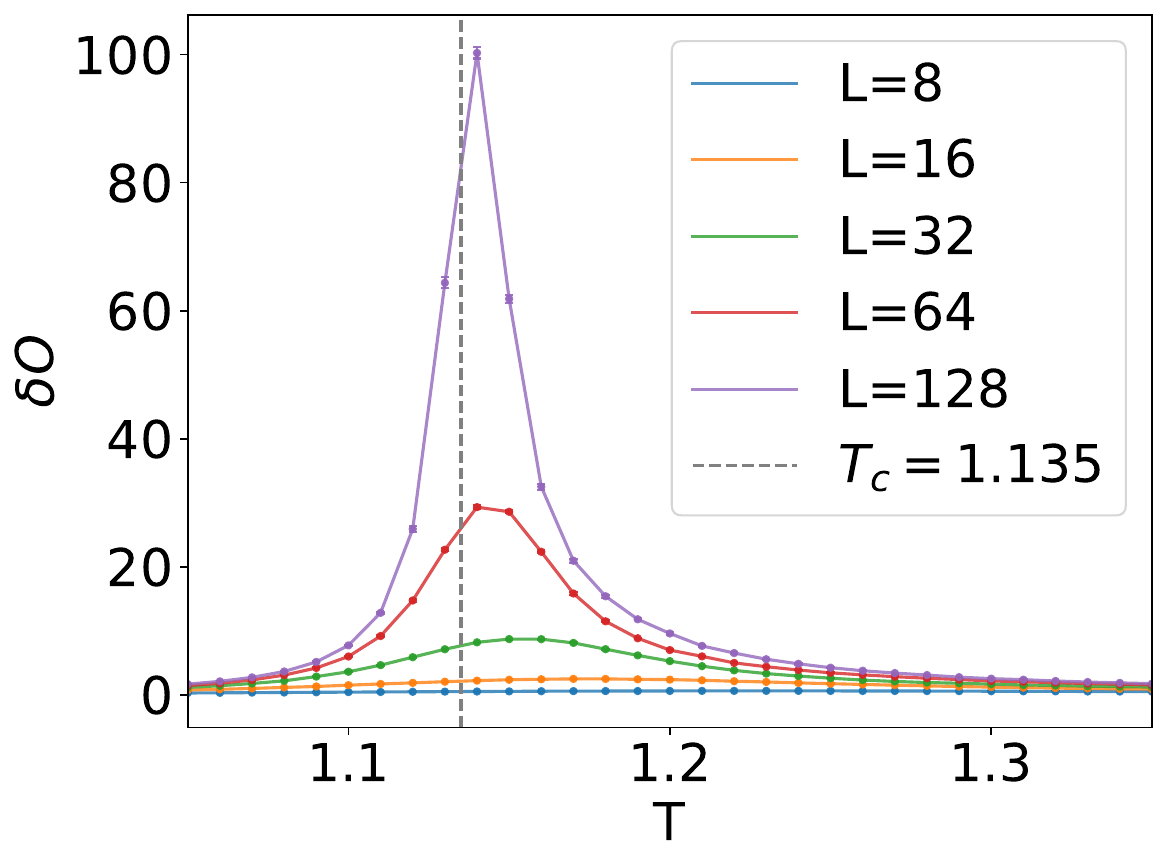}
\par
(b)
\end{minipage}
\begin{minipage}{0.32\textwidth}
\centering
\includegraphics[width=\textwidth]{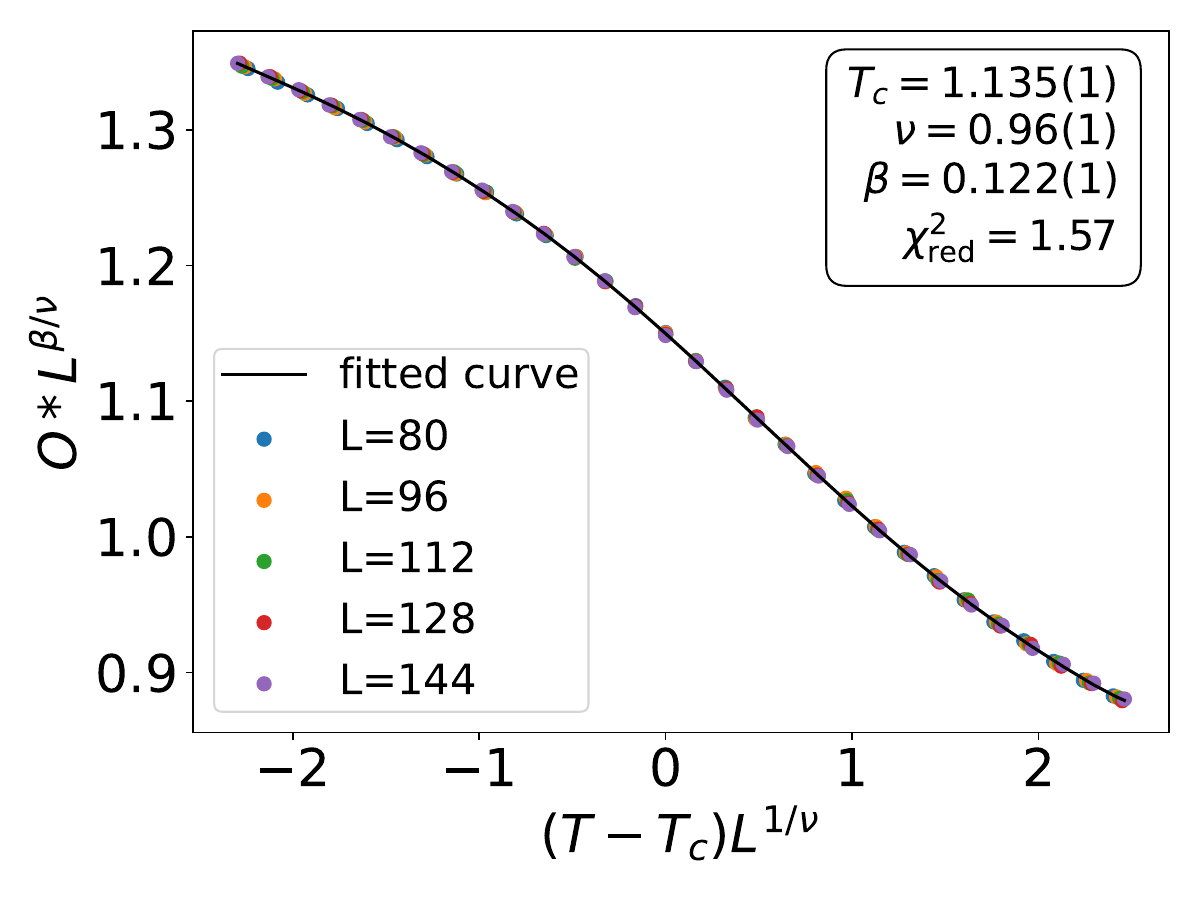}
\par
(c)
\end{minipage}
\caption{The temperature dependence of the maximal row correlation $O$ (a), and its fluctuations $\delta O$ (b) for the $q=2$ FM Potts model on the square lattice with various sizes. The vertical dashed lines in (a) and (b) mark the position of the exact critical temperature $T_c$ in the thermodynamic limit. The results of the data collapse with the optimized critical temperature and exponents with fluctuation range, including the value of the reduced chi-squared $\chi^2_{\text{red}}$, are shown in (c). In (b), error bars are shown for large lattice size ($L=128$ here). For all other cases, including the data for $O$ in panel (a), error bars are omitted due to their negligible magnitude, and the size of the symbols does not reflect the actual uncertainty.} 
\label{fig_square_fm_Q2}
\end{figure*}

\begin{figure*}
\centering
\begin{minipage}{0.32\textwidth}
\centering
\includegraphics[width=\textwidth]{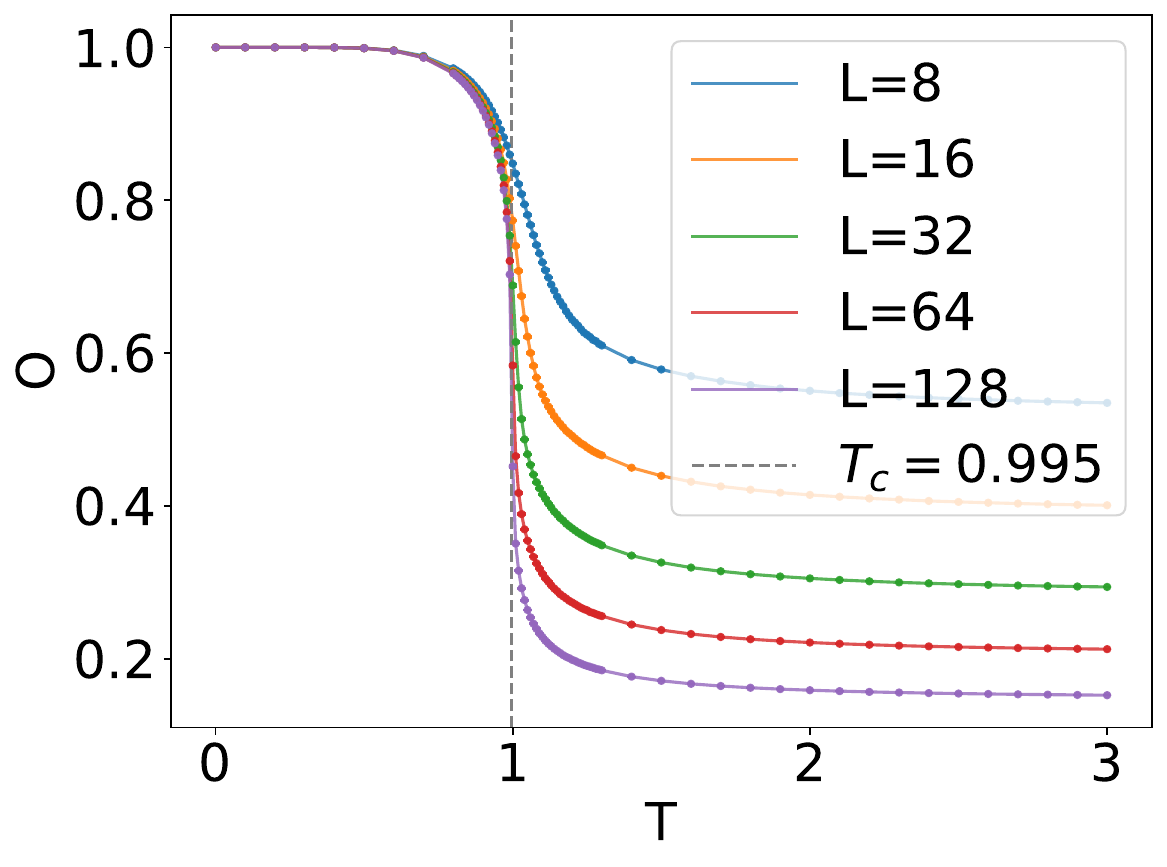}
\par
(a)
\end{minipage}
\begin{minipage}{0.32\textwidth}
\centering
\includegraphics[width=\textwidth]{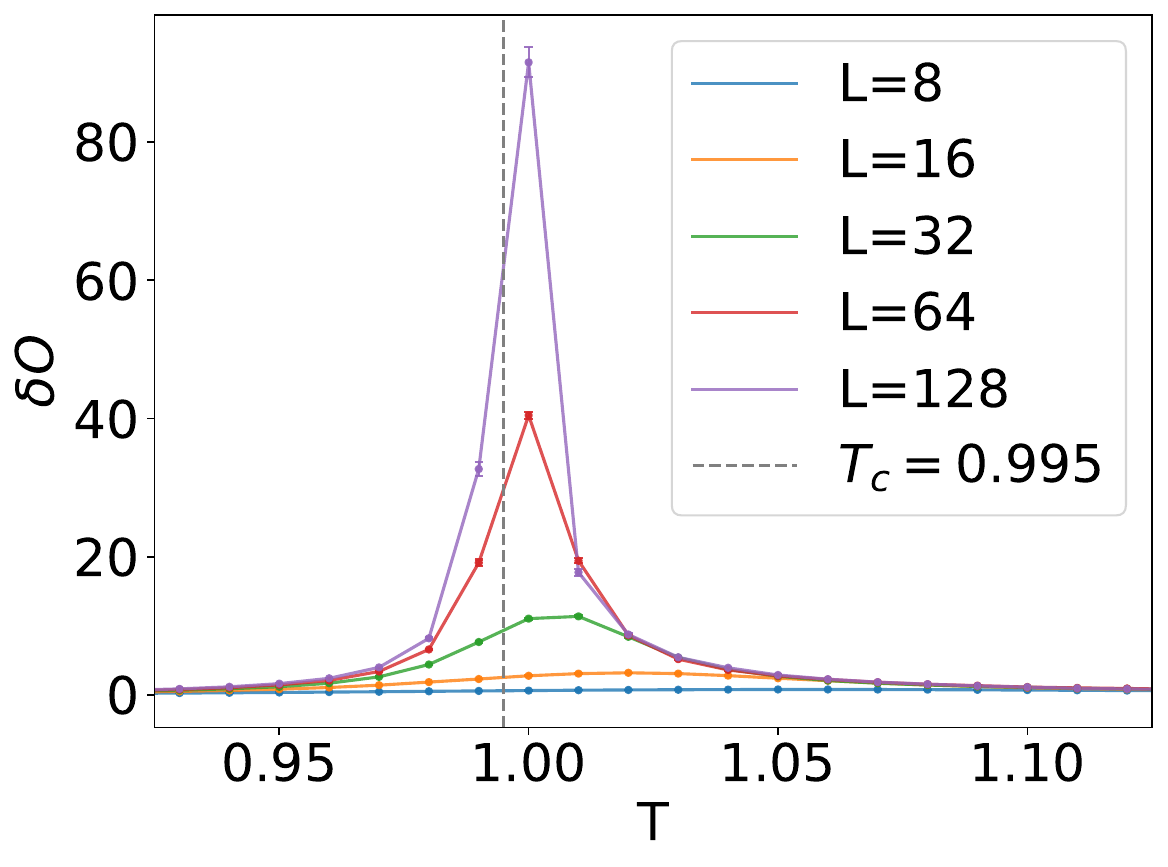}
\par
(b)
\end{minipage}
\begin{minipage}{0.32\textwidth}
\centering
\includegraphics[width=\textwidth]{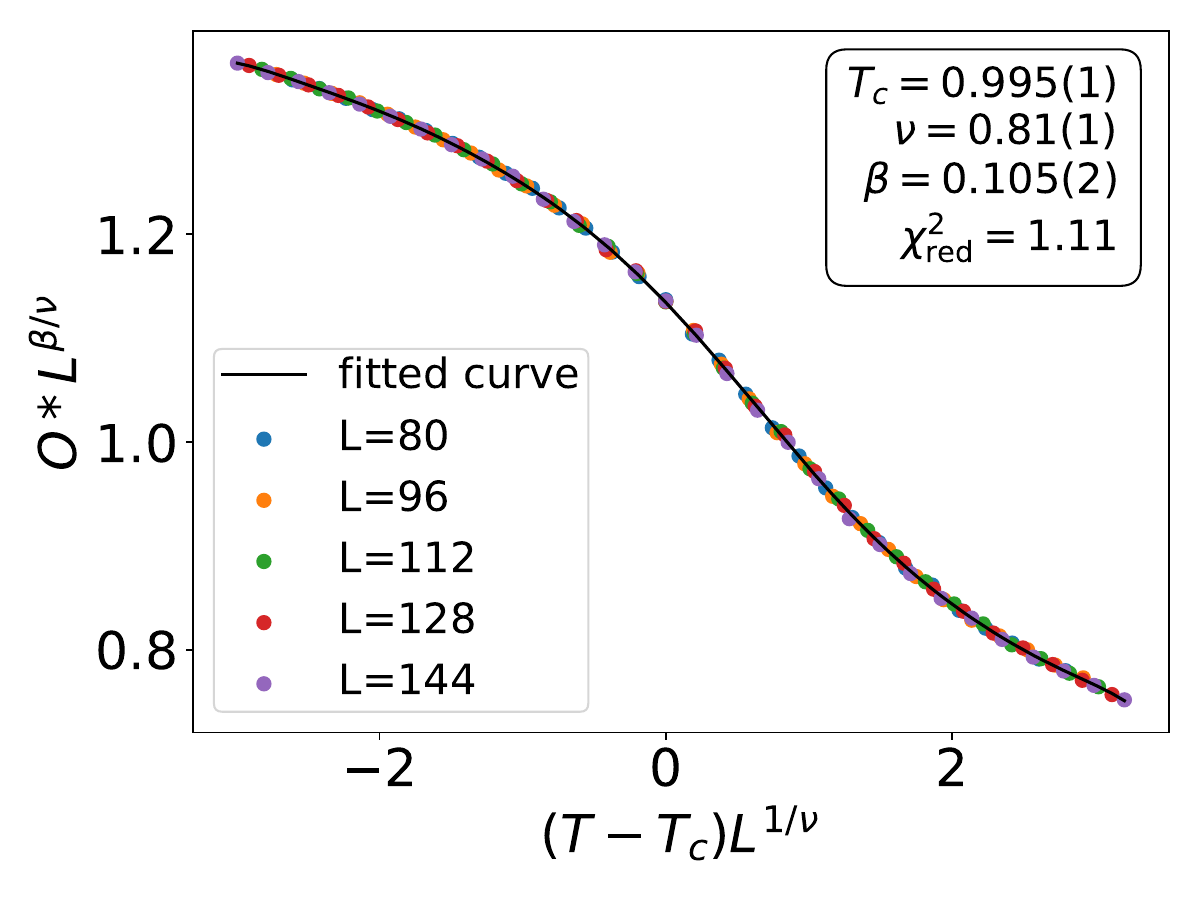}
\par
(c)
\end{minipage}
\caption{Same quantities as in Fig.~\ref{fig_square_fm_Q2}, but for the $q=3$ FM Potts model on the square lattice.}
\label{fig_square_fm_Q3}
\end{figure*}

To benchmark the effectiveness of the maximal row correlation $O$ as the order parameter in classical spin systems, we first perform the MC simulations on the $q=2$ and $q=3$ FM Potts models on the square lattice. The exact critical temperature in the thermodynamic limit of the $q$-state Potts model on the square lattice with isotropic FM coupling is known to be $T_c=J/\ln(1+\sqrt{q})$, where $J$ is the coupling strength~\cite{WuFY_1982,Baxter_1982a}. The critical exponents, relevant for our present study, are $\beta=1/8$, $\nu=1$ for the prestigious Ising case ($q=2$), and $\beta=1/9$, $\nu=5/6$ for the $q=3$ case~\cite{WuFY_1982,Baxter_1982a}. Our numerical MC results are presented in Figs.~\ref{fig_square_fm_Q2} and \ref{fig_square_fm_Q3}, respectively.

Figures~\ref{fig_square_fm_Q2}(a) and \ref{fig_square_fm_Q3}(a) show the temperature dependence of the order parameter $O$ for the two FM Potts models. The phase transition between the ordered phase and the paramagnetic phase can be clearly distinguished. In the ordered phase, as the temperature approaches to zero, the order parameter $O$ converges to unity. On the other hand, in the disordered (paramagnetic) phase, $O$ decreases monotonically as the temperature increases, and its high-temperature limit scales as $1/\sqrt{L}$, as discussed in Appendix~\ref{app_A}. The overall features of $O$ as a function of temperature and its lattice-size dependence are quite similar to those of magnetization. 

Figures~\ref{fig_square_fm_Q2}(b) and \ref{fig_square_fm_Q3}(b) show the temperature dependence of $\delta O$, the fluctuation of the order parameter. We see that as the lattice size $L$ increases, the peak becomes more pronounced and steadily approaches the critical temperature $T_c$ in the thermodynamic limit, marked by the vertical dashed line. We also note the similarity in the MC result of $\delta O$ and the conventional magnetic susceptibility $\chi$. 

The above observations suggest a close relationship in the FM Potts model between the maximal row correlation $O$ and the magnetization $M$, as well as between $\delta O$ and the susceptibility $\chi$. It might be not surprising, given that $O$ is proportional to the square root of the maximal eigenvalue of the row-row correlation matrix $M$, as discussed in the Methods section~\ref{sec_methods}. The correspondence between $O$ and $M$ (as well as between $\delta O$ and $\chi$) can be further reinforced by the calculations of the relevant critical exponents. As a demonstration, we apply the data collapse method (detailed in Sec.~\ref{sec_methods}) to analyze the MC data of $O$ around the phase transition point for various lattice sizes. This allow us to determine the optimal values of $T_c$, $\beta$ and $\nu$, along with their uncertainty estimates. The results are presented in Figs.~\ref{fig_square_fm_Q2}(c) and \ref{fig_square_fm_Q3}(c). 

We see that both for the $q=2$ and $3$ cases, the critical temperatures obtained numerically are in good agreement with the exact results. As for the critical exponents, when $q=2$ (2D Ising case), we have in our calculation $\nu=0.96$ and $\beta=0.122$, where the known exact values are $\nu=1$, $\beta=1/8=0.125$; when $q=3$, the numerical results are $\nu=0.81$ and $\beta=0.105$, where the exact values are known to be $\nu=5/6\approx 0.833$, $\beta=1/9\approx 0.111$. The deviations we suggest are mainly due to the finite size effect. In the FM cases, the accuracy of using $O$ as the order parameter is comparable to that of the conventional magnetization $M$ for the same lattice sizes. We note that a similar data collapse procedure has also been applied to $\delta O$ to extract the critical exponents $\gamma$, which produces consistent results that agree well with known exact values. More details can be found in Appendix~\ref{app_data_collapse}.

\subsection{$q=3$ AFM Potts model on the diced lattice}\label{subsec_AFM_diced}

As mentioned in the Introduction section~\ref{sec_intro}, the advantage of using the maximal row correlation $O$ as the order parameter becomes evident in spin systems where conventional magnetic orders are not readily identified. For demonstration, we discuss in this subsection the phase transition in the $q=3$ AFM Potts model on the diced lattice in terms of $O$ and $\delta O$.

\begin{figure}
\centering
\begin{minipage}{0.23\textwidth}
\centering
\includegraphics[width=\textwidth]{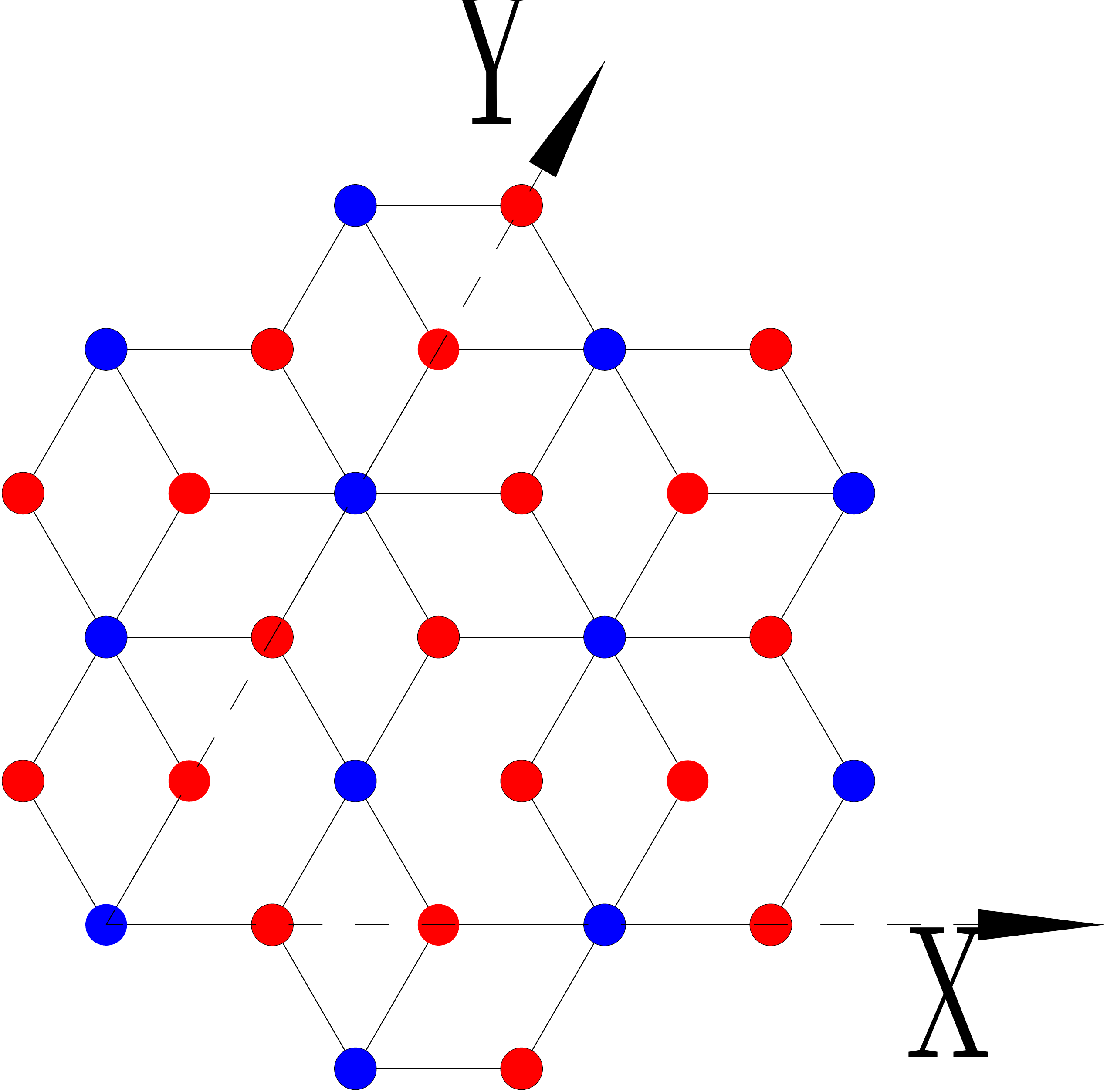}
\par\vspace{0.4em}
(a)
\end{minipage}
\hfill
\begin{minipage}{0.23\textwidth}
\centering
\includegraphics[width=\textwidth]{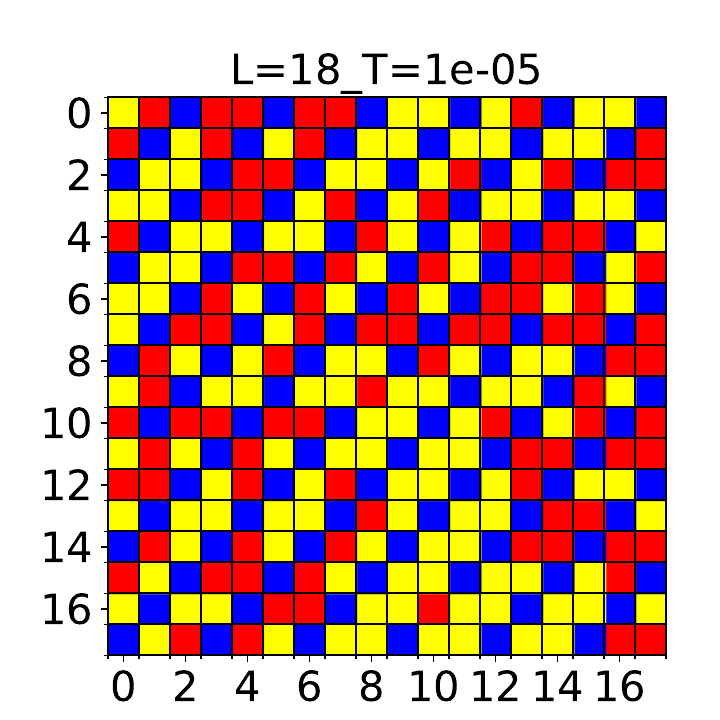}
\par
(b)
\end{minipage}
\caption{The diced lattice, as shown in (a), consists of a honeycomb sublattice (red circles, majority sites) and a triangular sublattice (blue circles, minority sites), where the latter is formed by sites located at the centers of the hexagons. Each blue circle is connected to six neighboring red circles, forming a bipartite structure. In order to map the spin configurations on the diced lattice into matrix form, an oblique rectilinear coordinate frame is used. In this coordinate frame, a blue circle appears after every two consecutive red circles along either the $x$ or $y$ direction. In (b) a snapshot of the spin configuration on $L=18$ diced lattice for the $q=3$ AFM Potts model is shown. It is taken from the MC sampling after the equilibration at the temperature $T=1.0\times 10^{-5}$. Three colors (red, blue and yellow) represent three spin values. A FM long-range order of one spin component (blue) is clearly observed on the triangular sublattice. In contrast, the spins on the honeycomb sublattice appear to randomly adopt the other two values (red and yellow).}
\label{fig_diced}
\end{figure}

The $q=3$ AFM Potts model is known to be a prototypical example which supports a phase with partial order in the low-temperature regime, i.e., a phase with a long-range order on one of its sublattices with minority sites, while another sublattice with majority sites remaining disordered~\cite{Kotecky_2008,ChenQN_2011} (see Fig.~\ref{fig_diced}). The phase transition from the disordered to the partially ordered phase is driven by the nontrivial entropy associated with the lattice's inherent geometric irregularity. 

\begin{figure*}
\centering
\begin{minipage}{0.32\textwidth}
\centering
\includegraphics[width=\textwidth]{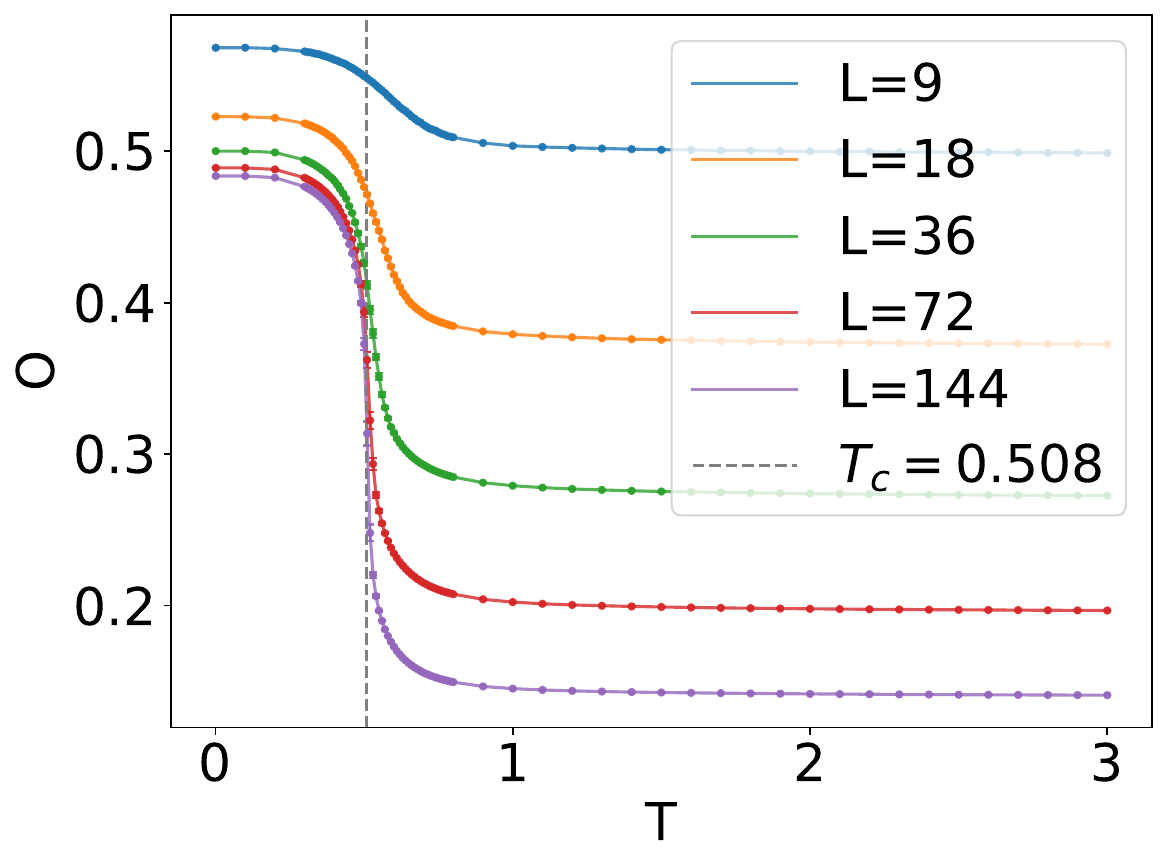}
\par
(a)
\end{minipage}
\begin{minipage}{0.32\textwidth}
\centering
\includegraphics[width=\textwidth]{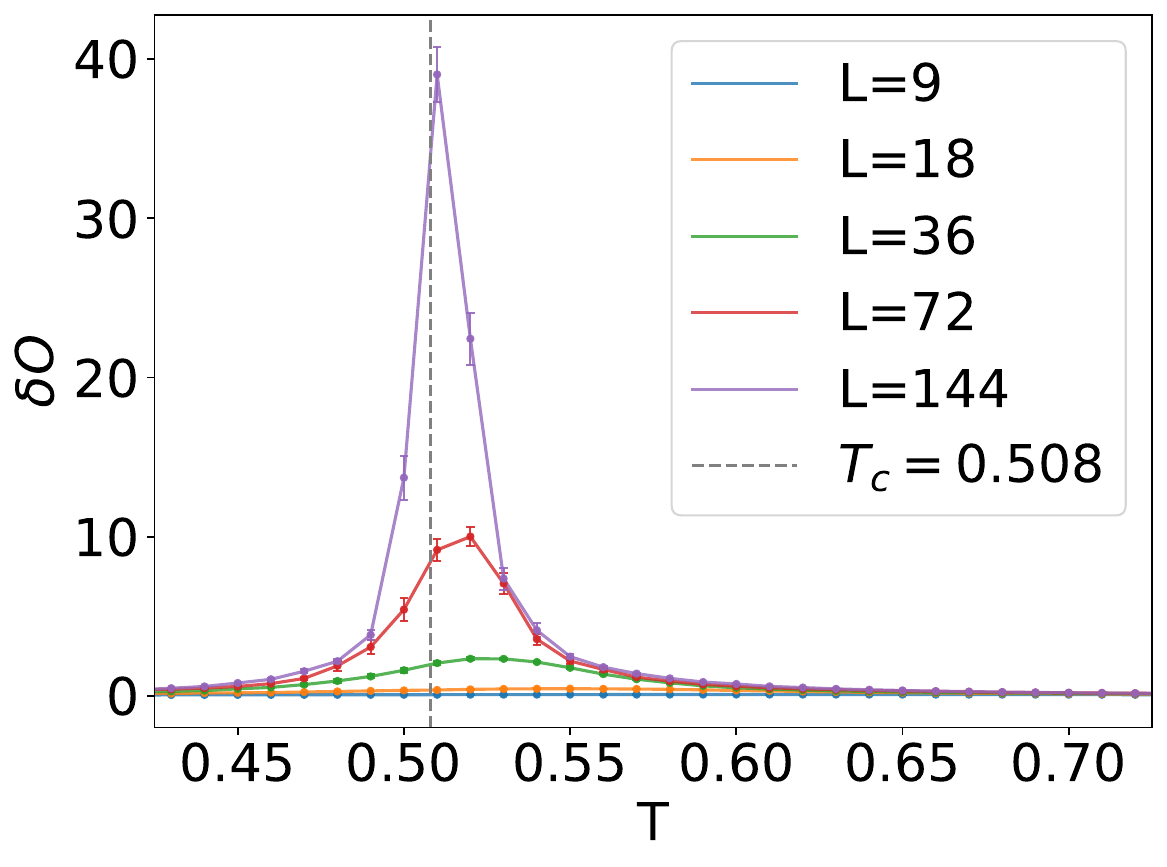}
\par
(b)
\end{minipage}
\begin{minipage}{0.32\textwidth}
\centering
\includegraphics[width=\textwidth]{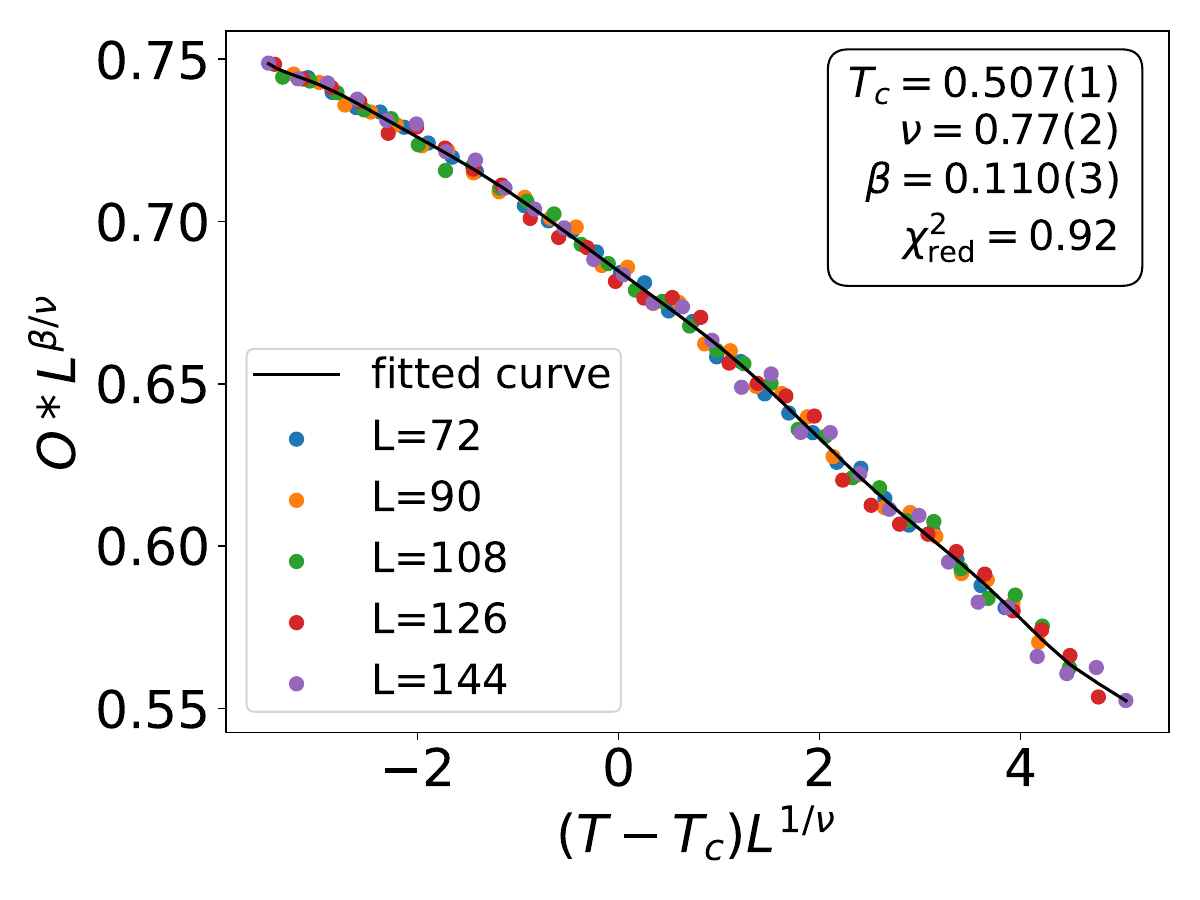}
\par
(c)
\end{minipage}
\caption{Same quantities as in Fig.~\ref{fig_square_fm_Q2}, but for the $q=3$ AFM Potts model on the diced lattice. The vertical dashed lines in (a) and (b) mark the position of the critical temperature $T_c$ obtained in Ref.~\cite{Kotecky_2008}.}
\label{fig_diced_afm_Q3}
\end{figure*}

In previous MC studies of the model's phase transition, the Binder-type ratio of a staggered magnetization was analyzed~\cite{Kotecky_2008}. In particular, the staggered magnetization, defined as $M_{\text{stagg}} = M_{\text{tri}} - M_{\text{hex}}$ , was introduced and used to characterize the magnetization difference between the triangular and honeycomb sublattices. In Ref.~\cite{ChenQN_2011}, the critical temperature was determined with high precision using a numerical tensor-based method, which identified a pronounced peak in the specific heat at $T_c=0.505(1)$. The critical temperature is in good agreement with the MC result $T_c=0.5075(1)$~\cite{Kotecky_2008}. In the following, analogous to the previous analysis of the FM cases, we present our MC results for the maximal row correlation $O$ and its fluctuation $\delta O$ in this model. To determined the critical temperature $T_c$ and the associated critical exponents $\nu$ and $\beta$, we apply the data collapse method to the MC data for $O$. The results are shown in Fig.~\ref{fig_diced_afm_Q3}.

From Figs.~\ref{fig_diced_afm_Q3}(a) and \ref{fig_diced_afm_Q3}(b), which display the temperature dependence of $O$ and $\delta O$, we see that despite the model considered here involving AFM interactions and exhibiting only partial order in its low-temperature phase, the qualitative behavior of the observables mirrors that of the FM cases presented in the previous subsection~\ref{subsec_FM}. The critical temperature, estimated from the peak position of $\delta O$ for the $L=144$ lattice, is already in close agreement with the reference values reported in Refs.~\cite{Kotecky_2008,ChenQN_2011}. These results provide compelling evidence for both the universality and the effectiveness of our approach. 

It is also worth noting that, in addition to the similarities between the AFM and FM cases, one notable difference arises in the low-temperature limit: as $T\to 0$, the order parameter $O$ for this model does not approach unity---as in the FM case---but instead converges to a nonzero value slightly below $1/2$ with apparent size dependence. This behavior, where the maximal row correlation approaches a stable value at low temperatures as the system size increases, reflects the presence of partial order in the ground state of the model, as discussed in detail in Appendix~\ref{app_A}. 

The critical temperature $T_c$ and the critical exponents $\nu$ and $\beta$, determined by the data collapse method, are shown in Fig.~\ref{fig_diced_afm_Q3}(c). The estimated critical temperature $T_c=0.507$ is in excellent agreement with the reference values. The critical exponents, $\nu=0.77$, with an uncertainty range of $[0.75,0.79]$ and $\beta=0.110$, are also consistent with those of the $q=3$ FM Potts model ($\nu=5/6\approx 0.833$, $\beta=1/9\approx 0.111$). These observations corroborate the conclusion in Ref.~\cite{Kotecky_2008} that the critical behavior of the model belongs to the universality class of the $q=3$ Potts ferromagnet. The data collapse of $\delta O$, along with the extracted critical exponents $\nu$ and $\gamma$ on this model, is presented in Appendix~\ref{app_data_collapse}.

One may note that in our data collapse analysis, which include the previous FM cases, the estimated accuracy of the critical exponent $\nu$ is somewhat lower than that of the exponent $\beta$, with the critical temperature $T_c$ exhibiting the highest accuracy. This can be attributed to the fact that $\nu$ is inherently more sensitive to statistical fluctuations and finite-size effects, especially in frustrated or irregular systems, as it controls the divergence of the correlation length---a non-local quantity---and enters the scaling functions through the rescaled temperature variable $\left(T-T_c\right)L^{1/\nu}$. In light of this established hierarchy of accuracy among the critical parameters, we regard the extracted value of $\nu$ as reasonable within the system sizes accessible in our study.

\subsection{$q=2$ and $q=3$ AFM Potts model on Union-Jack lattice}\label{subsec_AFM_unionjack}

\begin{figure}
\centering
\begin{minipage}{0.23\textwidth}
\par\vspace{2em}
\centering
\includegraphics[width=\textwidth]{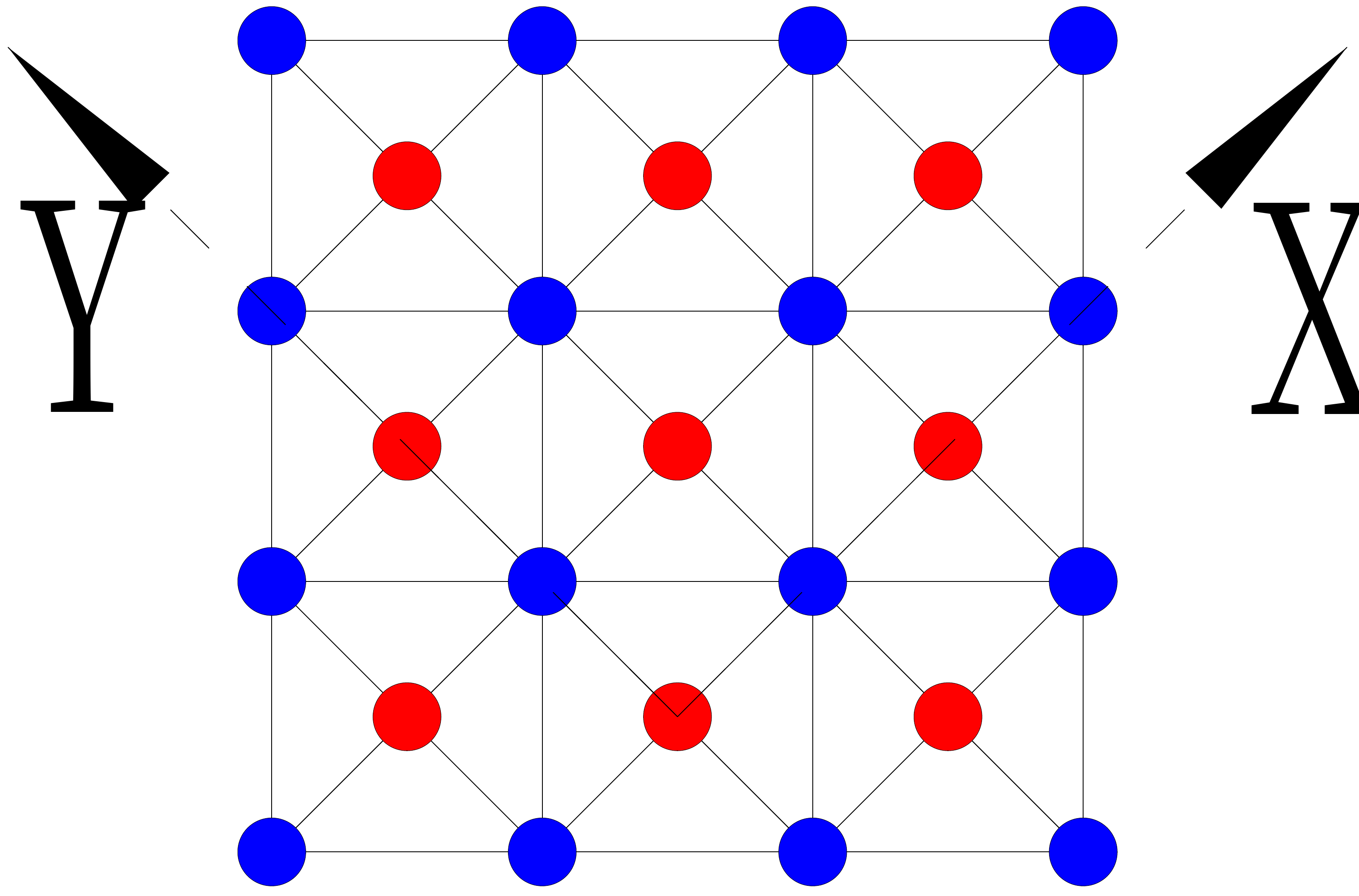}
\par\vspace{2em}
(a)
\end{minipage}
\hfill
\begin{minipage}{0.23\textwidth}
\centering
\includegraphics[width=\textwidth]{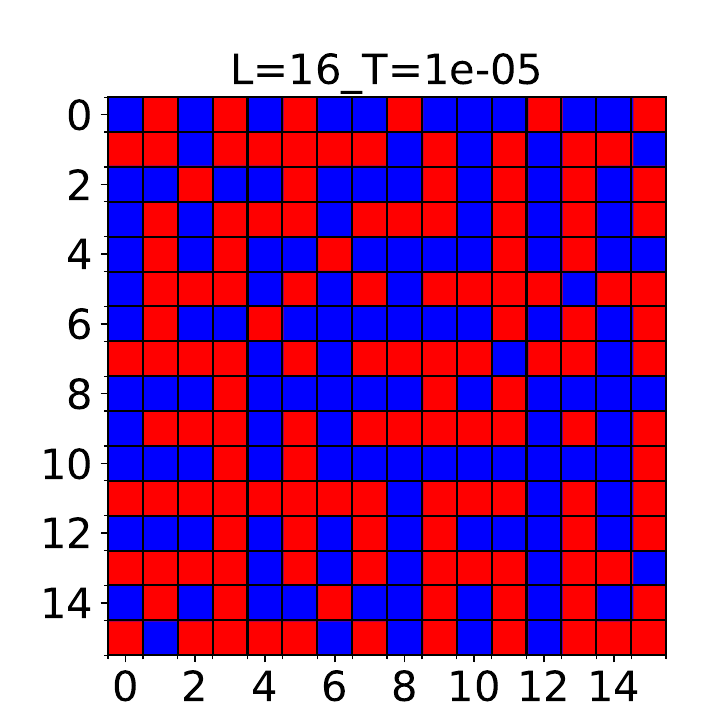}
\par
(b)
\end{minipage}
\caption{(a) The Union-Jack lattice has a tripartite structure, formed by inserting additional sites (red circles) at the centers of the plaquettes of a square lattice (blue circles). Each red site connects to its four nearest blue neighbors. The three sublattices have coordinate numbers $8$, $8$, and $4$, respectively, resulting in an average coordinate number $z=6$. To map the spin configuration on the lattice into matrix form, an orthogonal coordinate frame is employed. A snapshot of the spin configuration on $L=16$ Union-Jack lattice for the $q=2$ AFM Potts model is shown in (b), with two colors (red and blue) presenting two spin values. It is taken from the MC sampling after the equilibration at the temperature $T=1.0\times 10^{-5}$. We observe that due to the frustration, the absence of discernible order in the low-temperature configuration makes it challenging to identify an appropriate order parameter.}
\label{fig_union_jack}
\end{figure}

In this subsection, we study the $q=2$ and $q=3$ AFM Potts models on the Union-Jack lattice. Unlike the diced lattice---which is bipartite and free of frustration---the Union-Jack lattice is tripartite [Fig.~\ref{fig_union_jack}(a)], exhibiting frustration for $q=2$, whereas it remains unfrustrated for $q=3$.

For the $q=2$ case, exact analytical expressions of the free energy, as well as the spontaneous magnetization for both symmetric interaction and general anisotropic interactions have been obtained in Refs.~\cite{Vaks_1966} and ~\cite{WuFY_1987}, respectively. Entropy analysis reveals that, in the low-temperature limit, a partial AFM order emerges on two of its three sublattices (occupied by the blue circles in Fig.~\ref{fig_union_jack}(a)), while the third sublattice (red circles) remains disordered~\cite{Chen_2012}. However, unlike the frustration-free diced lattice, this partial order on the Union-Jack lattice is obscured by frustration, making it difficult to discern directly from the low-temperature configurations generated by the MC simulations (see Fig.~\ref{fig_union_jack}(b)). 

% Especially, for the uniform interaction, the critical temperature reads $T_c\approx 0.719348$~\cite{Vaks_1966}.
% It underlines directly the importance of the irregular nature of a 2D lattice in promoting a finite zero-temperature entropy and partial order in the ground state.

\begin{figure}
\centering
\begin{minipage}{0.23\textwidth}
\centering
\includegraphics[width=\textwidth]{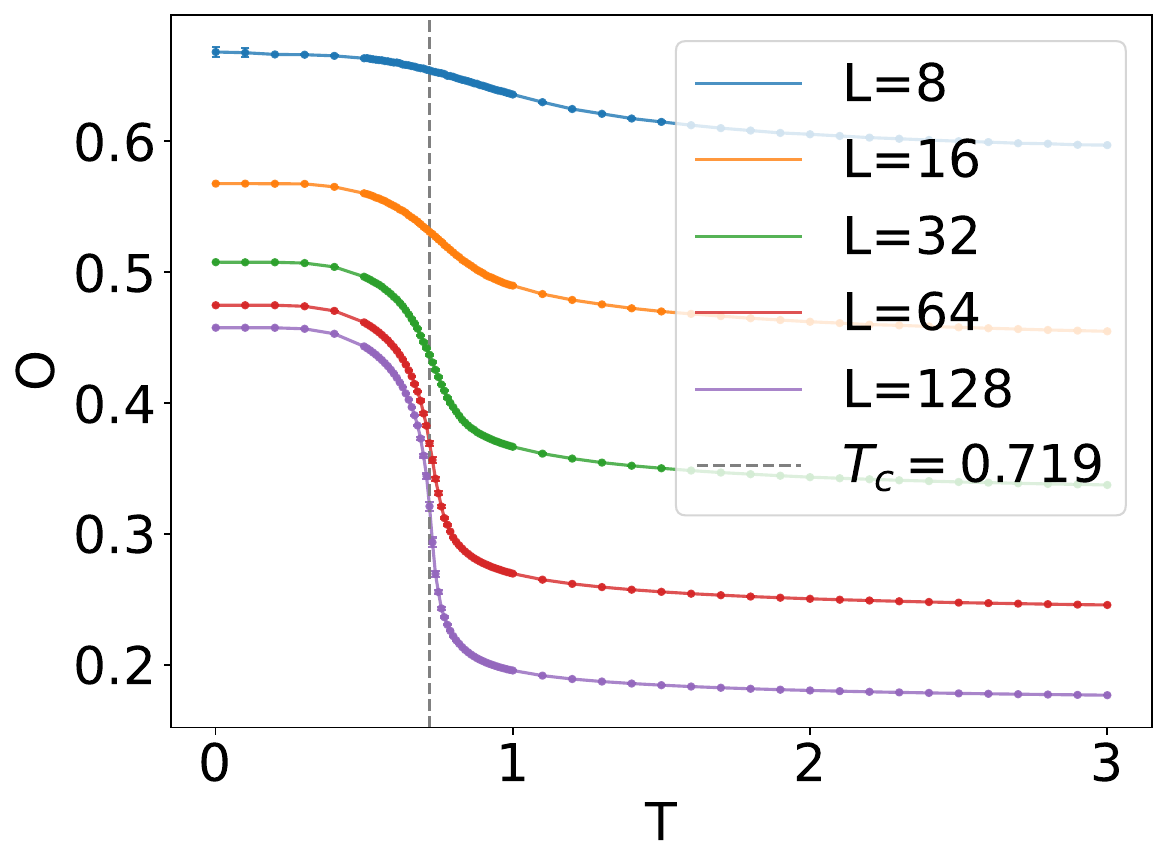}
\par
(a)
\end{minipage}
\begin{minipage}{0.23\textwidth}
\centering
\includegraphics[width=\textwidth]{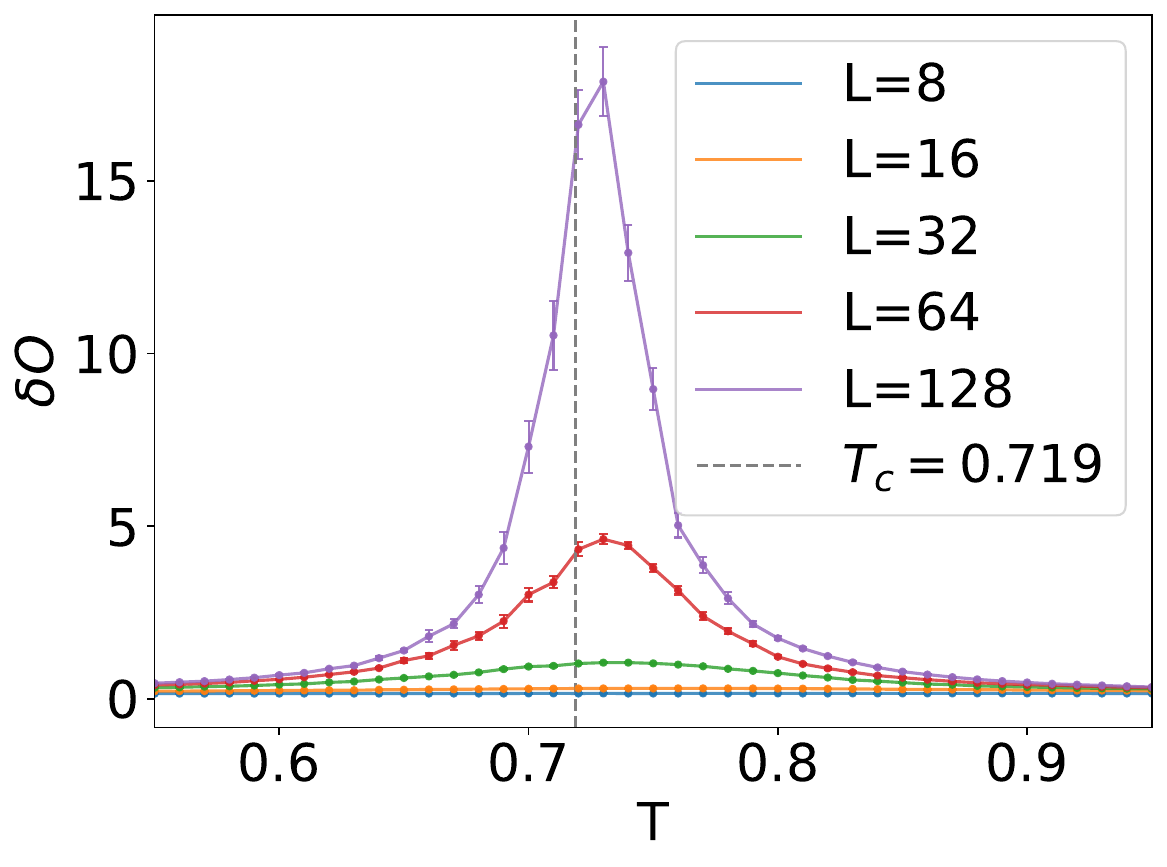}
\par
(b)
\end{minipage}
\caption{The temperature dependence of the maximal row correlation $O$ (a), and its fluctuations $\delta O$ (b) for the $q=2$ AFM Potts model on the Union-Jack lattice with various sizes. The vertical dashed lines in (a) and (b) mark the position of the exact critical temperature $T_c$ in the thermodynamic limit obtained in Ref.~\cite{Vaks_1966}.}
\label{fig_union_jack_afm_Q2}
\end{figure}

In contrast, the `hidden' partial order can be revealed in a straightforward manner by analyzing the maximal row correlation $O$ and its fluctuations $\delta O$. Figure~\ref{fig_union_jack_afm_Q2} presents the MC results for the temperature dependence of $O$ and $\delta O$. We see that at $T\to 0$, $O$ approaches a nonzero value (around $0.5$) as the system size $L$ increases, indicating the emergence of partial order in the low-temperature limit. Furthermore, the peak position of $\delta O$ for $L=128$ aligns closely with the exact critical temperature $T_c$, providing a reliable indicator of the phase transition.

\begin{figure}
\centering
\begin{minipage}{0.23\textwidth}
\centering
\includegraphics[width=\textwidth]{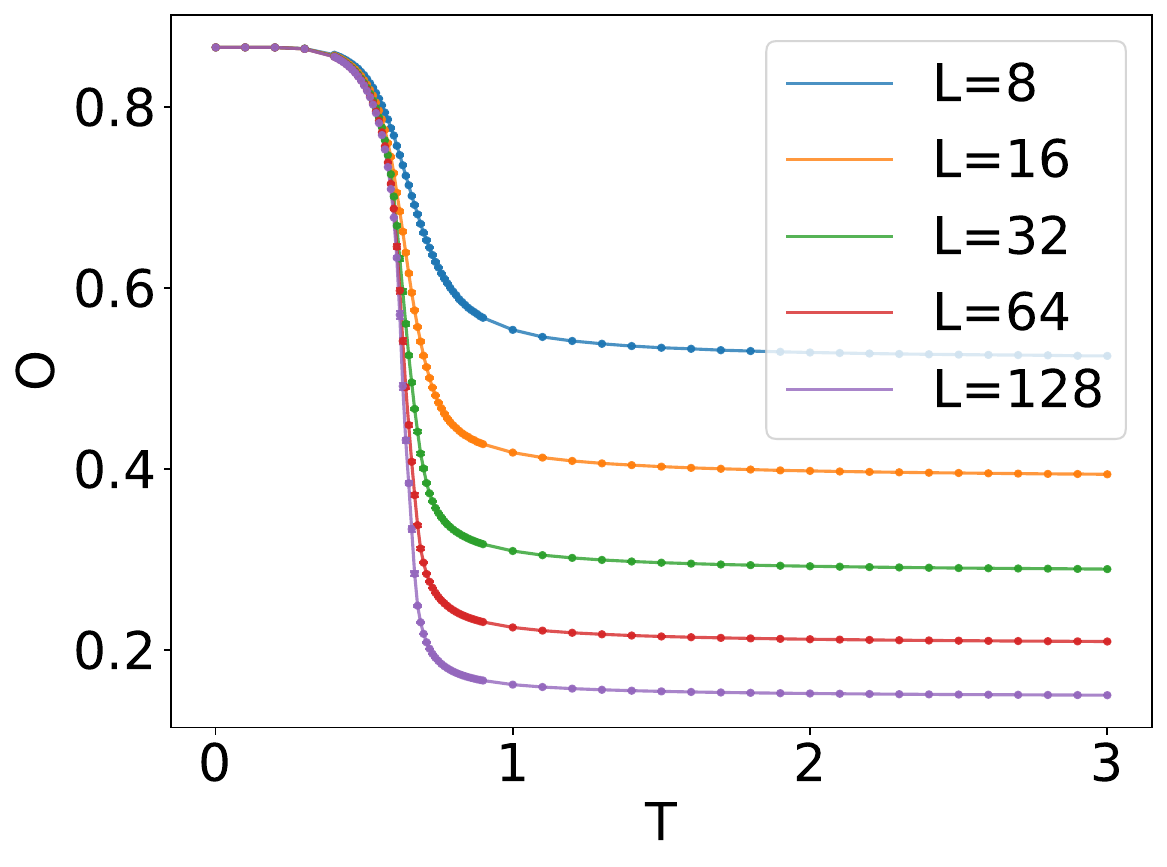}
\par
(a)
\end{minipage}
\begin{minipage}{0.23\textwidth}
\centering
\includegraphics[width=\textwidth]{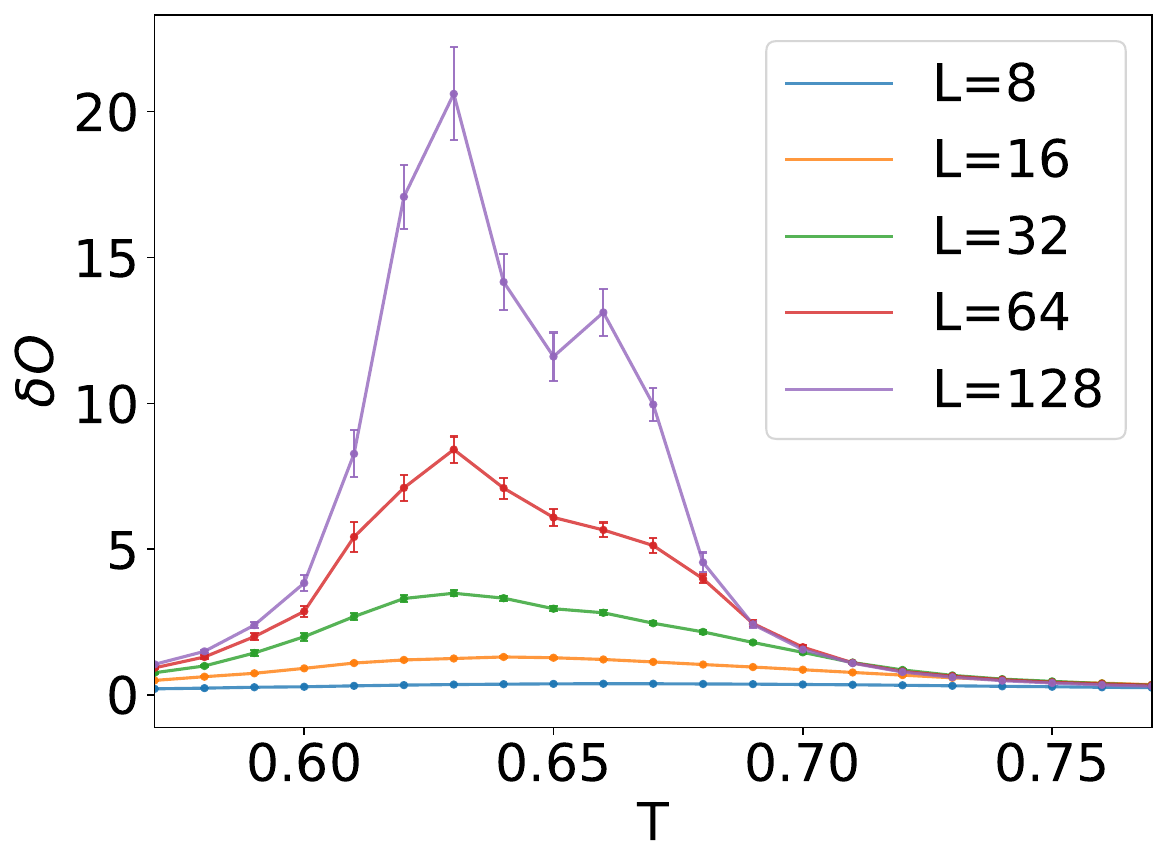}
\par
(b)
\end{minipage}
\caption{Same quantities as in Fig.~\ref{fig_union_jack_afm_Q2}, but for the $q=3$ case.}
\label{fig_union_jack_afm_Q3}
\end{figure}

For the case $q=3$, as noted at the beginning of this subsection, the system is unfrustrated. The ground-state configurations consist of each sublattice being occupied by spins of one of the three distinct values, such that each spin value appears on a different sublattice. Although the ground state is degenerate, the degeneracy is finite and does not scale with the system size; hence, there is no macroscopic degeneracy. Interestingly, as the temperature increases, the system undergoes two successive phase transitions: from an ordered phase to a partially ordered phase, and then to a fully disordered phase. The phenomenon was first reported in Ref.~\cite{Chen_2012}, based on tensor-network calculations of the specific heat. 

Our MC results on the maximal row correlation $O$ and its fluctuation $\delta O$ are presented in Fig.~\ref{fig_union_jack_afm_Q3}. We see that in contrast to the $q=2$ case, the asymptotic value of $O$ in the low-temperature limit exhibits only weak size dependence and attains a higher value (larger than $0.8$), signaling the presence of a well-ordered phase as $T\to 0$. Regarding $\delta O$, the pronounced double-peak structure is clearly observed for the largest system size studied, $L=128$. Notably, the positions of these peaks align closely with those identified from the specific heat, reinforcing the conclusion of two successive phase transitions, as originally reported in Ref.~\cite{Chen_2012}.

\section{Summary and discussion}\label{sec_summary}

In this study, we demonstrate that the quantity of the maximal row correlation $O$, whose magnitude is proportional to the length of the longest vectors obtainable by linear superposition of the row vectors in the spin configuration matrix, captures a key aspect of the intrinsic spin correlations within a given configuration. MC simulations on the Potts models confirm the effectiveness of $O$ as an order parameter. Specifically, we find that $O$ approaches a finite value as $T\to 0$ whenever some degree of spin ordering (either partial or full) is present, regardless of the underlying lattice geometry, interaction type (FM or AFM), or the presence of frustration. In the high-temperature limit where spin order is destroyed, $O$ vanishes consistently. The onset of this critical behavior in $O$---or, more sharply, the peak position of its fluctuations $\delta O$---can be used to accurately identify the critical temperature of the phase transition. These results establish a general and robust framework for investigating phase transitions in classical spin systems, with $O$ serving as a versatile and broadly applicable order parameter.

We further show that the critical behavior of $O$, especially the associated critical exponents, closely parallels that of conventional magnetization---whether full, partial, or staggered---which typically requires careful definition in partially ordered or frustrated systems. As anticipated, the fluctuation $\delta O$ appears to correspond to a susceptibility-like quantity. This correspondence is quantitatively confirmed for the FM cases and the AFM case on the diced lattice [see Appendix~\ref{app_data_collapse}]. However, accurately extracting the associated critical exponents in systems with irregular lattices or frustration requires more extensive simulations on larger system sizes. We defer this investigation to future work.

We conclude this section by speculating potential applications of the order parameter $O$ introduced in this study. As discussed, $O$ is expected to serve as a unified order parameter applicable to a broad class of spin systems. Notably, since $O$ is derived from the row correlation matrix---specifically, it is proportional to the square root of the matrix's largest eigenvalue---its inherently nonlocal character may offer particular utility in identifying exotic phases that lack local order but exhibit nontrivial global structures. 

Moreover, we note that this technique bears resemblance to PCA, a widely used feature extraction technique in machine learning that has been applied to detect and characterize phase transitions in various models~\cite{WangLei_2016,ZhaiHui_2017,Hu_2017,ChenXS_2019,Haldar_2024,Tirelli_2022}. However, as emphasized in Sec.~\ref{subsec_methods_O}, the row correlation matrix $M$ differs from the covariance matrices employed in PCA, since it is constructed directly from individual spin configurations. As a result, the order parameter $O$ is closely tied to spin correlations and magnetic orders, offering clear physical interpretability. This framework also enables the extraction of the associated critical exponents through statistical and scaling analysis of $O$ and its fluctuation $\delta O$. Because our approach operates directly on MC-generating configurations without requiring prior knowledge of the system’s symmetry or structure, it may offer novel insights into the mechanisms underlying machine learning–based phase classification~\cite{Santos_2021a, Santos_2021b, Sale_2022}. Integrating this method with modern machine learning techniques thus presents a promising avenue for future research.

\section*{Acknowledgment}
We thank Li-Ping Yang for helpful discussions. We acknowledge support from the National Natural Science Foundation of China (NSFC; Grants No. 12174168 and No. 12247101). 

\section*{Data Availability}

The data that support the findings of this article are openly available~\cite{Codes}.

\appendix

\section{The maximal row correlation in various limiting cases}\label{app_A}

In this appendix, we present a qualitative discussion of the behavior of the maximal row correlation in various limiting cases, including the completely disordered phase at high temperature, and the fully or partially ordered phases at low temperature. The latter is particularly relevant to systems on irregular lattices, such as the diced and Union-Jack lattices discussed in Section~\ref{sec_results}.

\subsection{High temperature limit}

In the high-temperature limit, the spin orientations are completely disordered, and the inter-row correlations vanish on average. Consequently, the row correlation matrix $M$, which has been defined in the Introduction Section~\ref{sec_intro}, takes the form of a diagonal matrix on average: 
\begin{equation}
M =
\begin{bmatrix}
L & 0 & \cdots & 0 \\
0 & L & \cdots & 0 \\
\vdots & \vdots & \ddots & \vdots \\
0 & 0 & \cdots & L
\end{bmatrix}.
\end{equation}
Here and throughout the following discussion in this appendix, we assume without loss of generality that the configuration matrix $A$ is square, i.e., both the number of rows and the number of columns is $L$. 

It is easy to see that the maximal eigenvalue of $M$ is simply $L$, and by definition, the maximal row correlation reads $O=\sqrt{L}/L=1/\sqrt{L}$. Thus in the disordered case, the order parameter $O$ vanishes asymptotically as $1/\sqrt{L}$ with increasing lattice size.

\subsection{Low temperature limit}

In this subsection, we demonstrate that for ordered states, the maximal row correlation $O$ remains finite in the thermodynamic limit.

\subsubsection{Full long-range order}

For simplicity, here we focus on the fully polarized configuration, for which the row correlation matrix $M$ is given by
\begin{equation}
M=
\begin{bmatrix}
L & L & \cdots & L\\
L & L & \cdots & L\\
\vdots & \vdots & \ddots & \vdots\\
L & L & \cdots & L
\end{bmatrix},
\label{eq_M_ordered}
\end{equation}
where $L$ is the system size. 

The only nonzero eigenvalue of $M$ is $L^2$, with the corresponding eigenvector $\mathbf{v}=(1,1,\ldots,1)^T$. This yields a maximal row correlation of unity, since $O=\sqrt{L^2}/L=1$. We thus infer that under the definition $O=\sqrt{\lambda_M}/L$, the maximal value of $O$ cannot exceed one, and that this upper bound is attained only in completely ordered configurations.

\subsubsection{Partial order in the $q=3$ AFM Potts model on the diced lattice}

In the low-temperature limit, partial order in the $q=3$ AFM Potts model on the diced lattice emerges on the minority sublattice (blue circles in Fig.~\ref{fig_diced}(a)), where spins adopt a fixed value. In contrast, spins on the majority sublattice (red circles forming a honeycomb sublattice) fluctuate between the remaining two values. As a result, on average the row correlation matrix $M$ under the oblique rectilinear coordinate frame (see Fig.~\ref{fig_diced}(b)) takes the form:
\begin{equation}
\begin{bmatrix}
L & -L/4 & -L/4 & L/2 & -L/4 & \ldots \\
-L/4 & L & -L/4 & -L/4 & L/2 & \ldots \\
-L/4 & -L/4 & L & -L/4 & -L/4 & \ldots \\
L/2 & -L/4 & -L/4 & L & -L/4 & \ldots \\
% -L/4 & L/2 & -L/4 & -L/4 & L & \ldots \\
\vdots & \vdots & \vdots & \vdots & \ddots& \vdots \\
% -L/4 & -L/4 & L/2 & -L/4 & -L/4 & \cdots & L
\ldots & \ldots & L/2 & -L/4 & -L/4 & L
\end{bmatrix}.
\label{eq_M_for_diced}
\end{equation}
More concisely, the matrix element of $M$ can be written as
\begin{equation}
m_{ij}=\left\{\begin{array}{ll}
L, & i=j; \\
-L/4, & \abs{i-j}\equiv 1,2\pmod{3}; \\
L/2, & \abs{i-j}\equiv 0\pmod{3} \text{ and }\ i\neq j.
\end{array}\right.
\end{equation}

The expectation values of the matrix element $m_{ij}$ are easy to calculate since it is simply the inner product of the $i$-th and $j$-th rows of the spin configuration matrix $A$. To calculate the inner product, we need to specify the product between two spin variables, which is given by Eq.~(\ref{eq_spinproduct}) for the Potts model. For the diced lattice, when the row indices satisfy $\abs{i-j}\equiv 1,2\pmod{3}$, a straightforward counting shows that among the $L$ spin pair products appearing in the inner product expression (see Eq.~(\ref{eq_innerproduct})), two-thirds involve a product between an ordered site (with a fixed spin) and a random site (with a random spin), each contributing an expectation value of $-1/2$. The remaining one-third are products between two random sites, yielding an expectation value $\frac{1}{4}(2\times 1+2\times(-\frac{1}{2}))=1/4$. Thus in this situation we have
\begin{equation*}
m_{ij}=\frac{2}{3}L\times\left(-\frac{1}{2}\right)+\frac{L}{3}\times\frac{1}{4}=-\frac{L}{4},
\end{equation*}
Similar analysis is applied to the case $\abs{i-j}\equiv 0\pmod{3}\ \&\ i\neq j$, where there are one-third products between two ordered sites, and two-thirds between two random sites. It yields
\begin{equation*}
m_{ij}=\frac{L}{3}+\frac{2}{3}L\times\frac{1}{4}=\frac{L}{2}.
\end{equation*}

\begin{figure}
\centering
\includegraphics[width=0.45\textwidth]{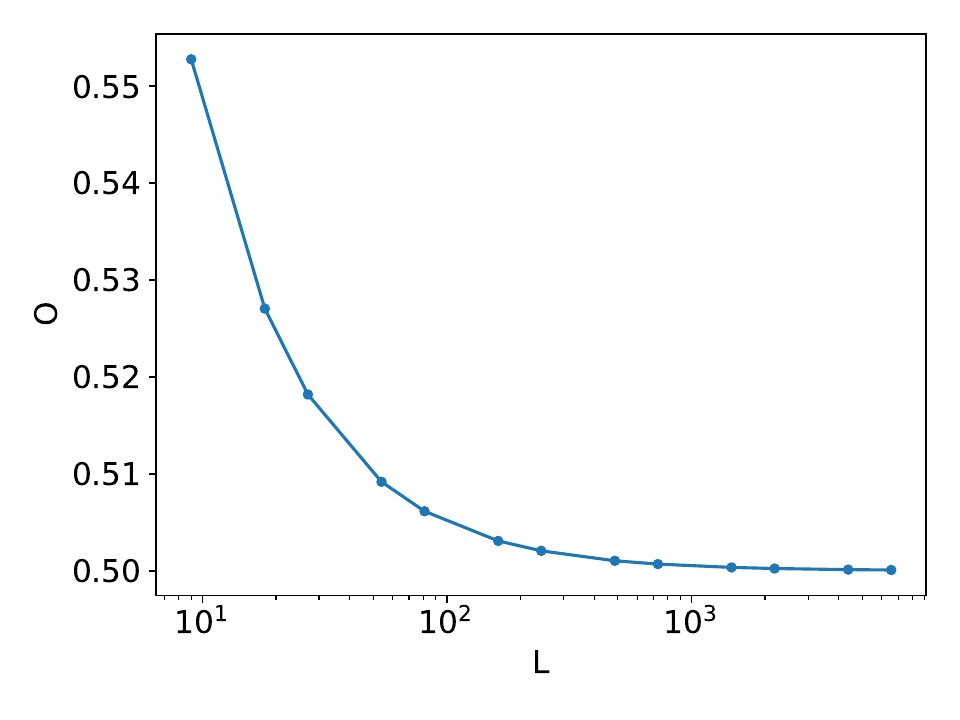}
\caption{Numerical results for the maximal row correlation $O$ as a function of $L$, computed from the row correlation matrix $M$ in Eq.~(\ref{eq_M_for_diced}).}
\label{fig_matrix}
\end{figure}

The maximal row correlation $O$ is defined as $O=\sqrt{\lambda_{\mathrm{M}}}/L$, where $\lambda_{\mathrm{M}}$ denotes the largest eigenvalue of the matrix $M$. We did not pursue an analytic expression for $\lambda_{\mathrm{M}}$; however, one can be easily convinced that for the matrix $M$ of dimension $L$ in the form of Eq.~(\ref{eq_M_for_diced}), the largest eigenvalue $\lambda_{\mathrm{M}}$ should scales as $\mathcal{O}(L^2)$. Numerical results for $O$ as a function of $L$, shown in Fig.~\ref{fig_matrix}, indicates that the leading term in $\lambda_{\mathrm{M}}$ is equal to $L^2/4$, yielding $\lim_{L\to\infty}O=1/2$. We observe that, in contrast to fully ordered cases---such as the FM models where $O$ approaches unity in the low-temperature limit with only weak size dependence---the asymptotic value of $O$ in systems exhibiting partial order typically converges to a nonzero value (usually less than one) with apparent size dependence.

The behavior of $O$, as obtained from a simple analysis here based on the picture of partial order/disorder on the triangular/honeycomb sublattices in the low-temperature limit, is consistent with the MC results in Fig.~\ref{fig_diced_afm_Q3}(a). These results show that $O$ approaches a nonzero value from above as the system size $L$ increases. The fact that the MC data converge to a value slightly below $1/2$ is attributed to the presence of a portion of defect configurations on the ground-state manifold (see also Fig.~\ref{fig_diced}(b)). Although the number of such configurations is not small, their statistical weight is subdominant compared to that of the perfectly order states. These defect configurations---which also have the same ground-state energy---slightly lower the asymptotic value of $O$ from $1/2$.

\section{Data collapse results for $\delta O$}\label{app_data_collapse}

\begin{figure*}
\centering
\begin{minipage}{0.32\textwidth}
\centering
\includegraphics[width=\textwidth]{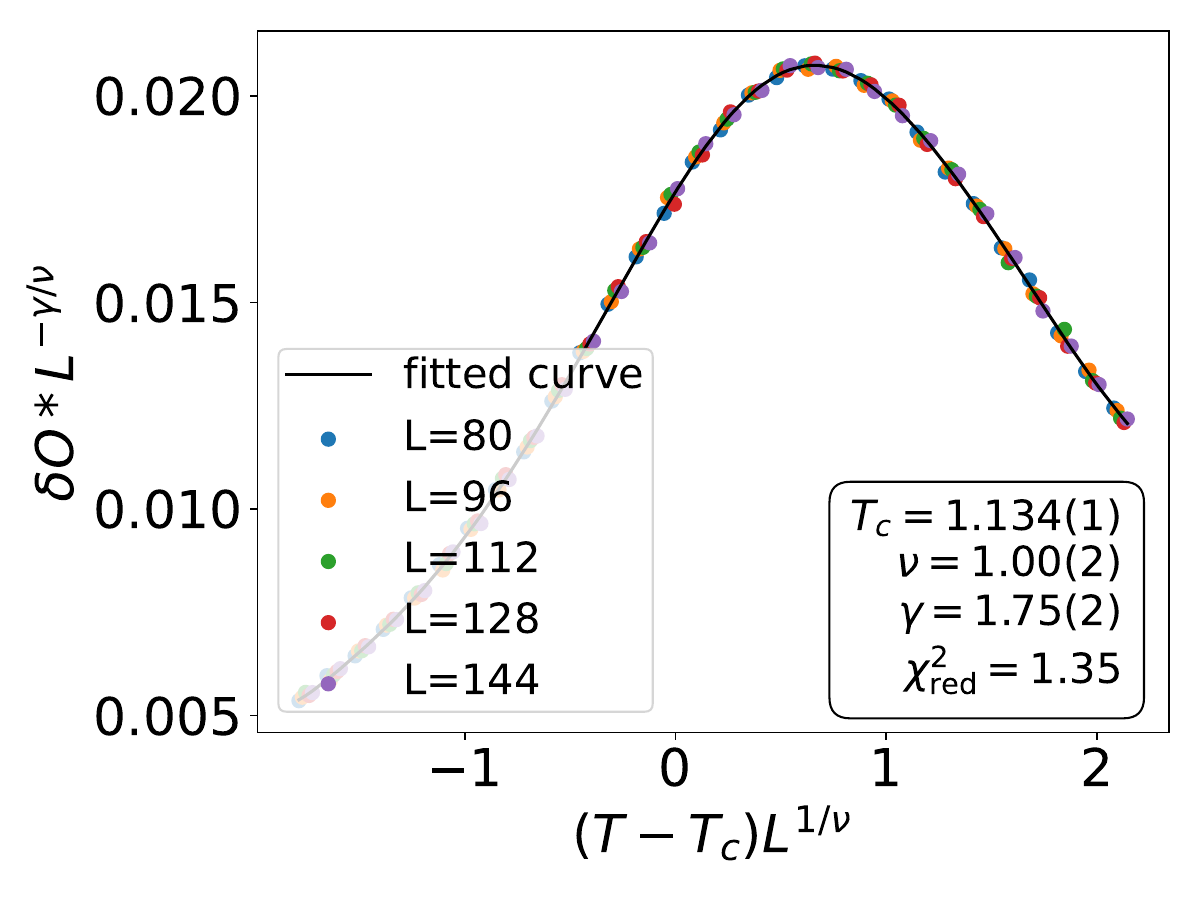}
\par
\textbf{(a)} 
\end{minipage}
\begin{minipage}{0.32\textwidth}
\centering
\includegraphics[width=\textwidth]{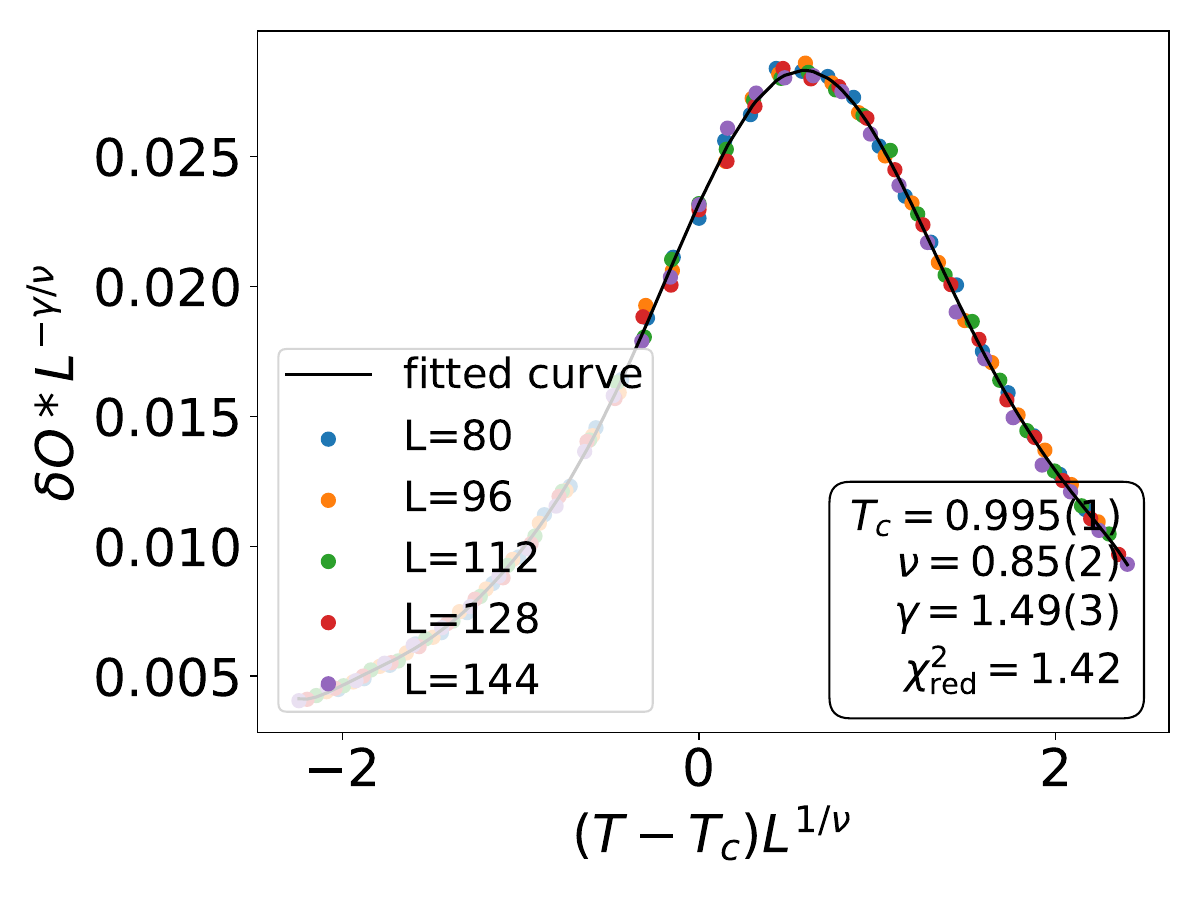}
\par
\textbf{(b)} 
\end{minipage}
\begin{minipage}{0.32\textwidth}
\centering
\includegraphics[width=\textwidth]{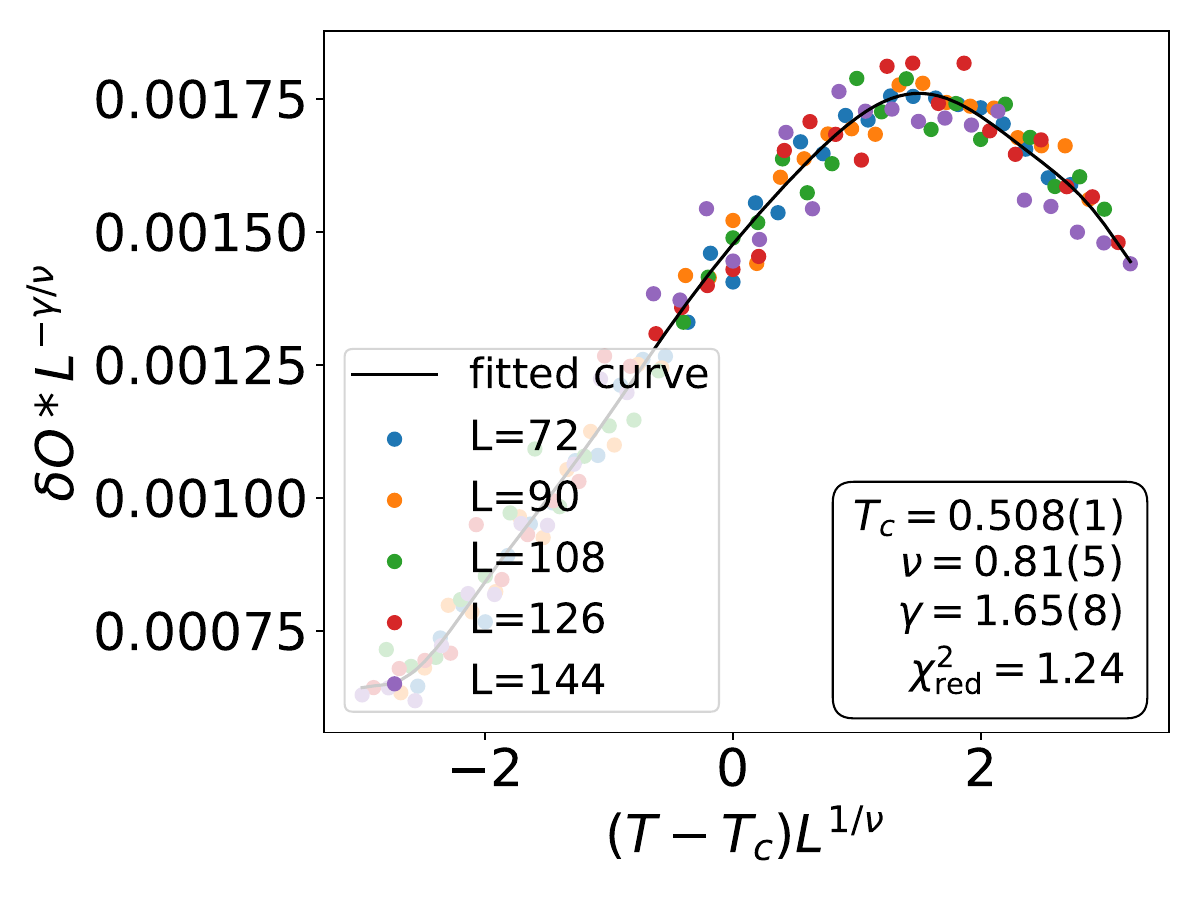}
\par
\textbf{(c)} 
\end{minipage}
\caption{Results of the data collapse for the fluctuations of the maximal row correlation $\delta O$ across different models: (a) the $q=2$ FM Potts model on the square lattice; (b) the $q=3$ FM Potts model on the square lattice; (c) the $q=3$ AFM Potts model on the diced lattice.}
\label{fig_data_collapse_nu}
\end{figure*}

From the previous discussion in Sec.~\ref{sec_results}, we have learned that the critical behavior of the maximal row correlation $O$ is governed by the critical exponent $\beta$, traditionally associated with the magnetization $M$. In this appendix, we further show that, analogously, the scaling behavior of its fluctuations $\delta O$ near the critical temperature can be used to extract the standard critical exponent $\gamma$. As a demonstration, Fig.~\ref{fig_data_collapse_nu} presents the results of the data collapse of $\delta O$ for the $q=2$ and $q=3$ FM Potts models on the square lattice, and the $q=3$ AFM Potts model on the diced lattice.

As shown in Fig.~\ref{fig_data_collapse_nu}, we see that the data collapse for FM interactions is highly satisfactory for both $q=2$ (Ising) and $q=3$ cases. For $q=2$, the collapse aligns well with the exact critical exponents $\nu=1$, and $\gamma=7/4=1.75$~\cite{Baxter_1982a} [Fig.~\ref{fig_data_collapse_nu}(a)]. For $q=3$, the known exact values are $\nu=5/6\approx 0.833$ and $\gamma=13/9\approx 1.444$~\cite{WuFY_1982,Baxter_1982a}; our numerical results yield $\nu=0.85(2)$, $\gamma=1.49(2)$ [Fig.~\ref{fig_data_collapse_nu}(b), which are in good agreement within numerical uncertainty.

In contrast, for the $q=3$ Potts model with AFM interactions on the diced lattice, the data collapse is notably less precise when using the same set of system sizes as in the $q=2$ and $q=3$ FM Potts models. The optimal values of $T_c$, $\nu$, and $\gamma$ yield a visible more scattered distribution of data points [Fig.~\ref{fig_data_collapse_nu}(c)], resulting in a comparatively large numerical uncertainties associated with the estimates of the critical exponents $\nu$ and $\gamma$. This reduced convergence may be attributed to the geometric irregularity of the diced lattice and the presence of partial ordering fluctuations. Achieving comparable accuracy would likely require simulations on larger system sizes. Nevertheless, the estimated values of $\nu$ and $\gamma$ remain consistent with the conclusion of Ref.~\cite{Kotecky_2008}, which suggests that the universality class of the $q=3$ AFM Potts model on the diced lattice is equivalent to that of the $q=3$ Potts ferromagnet.

\section{Higher dimensional systems}\label{app_highD}

\begin{figure}
\centering
\begin{minipage}{0.23\textwidth}
\centering
\includegraphics[width=\textwidth]{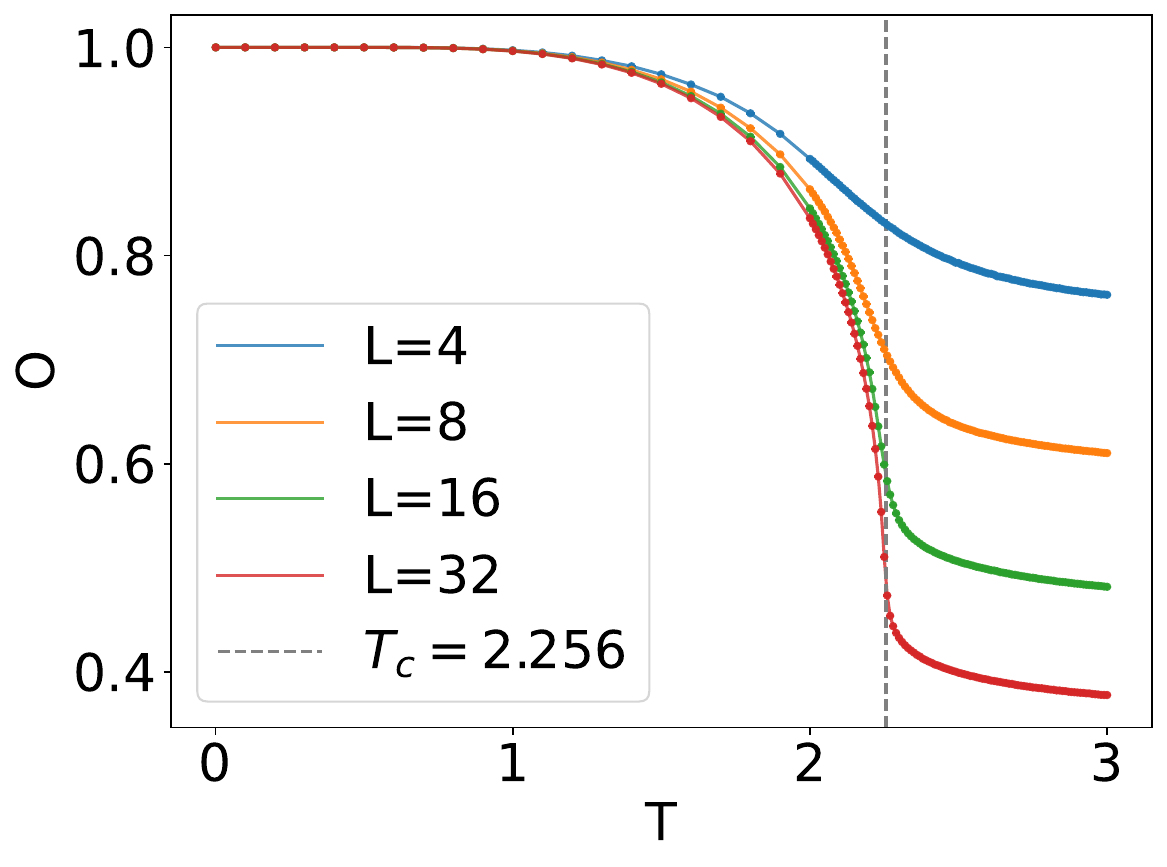} 
\par
(a)
\end{minipage}
\hfill
\begin{minipage}{0.23\textwidth}
\centering
\includegraphics[width=\textwidth]{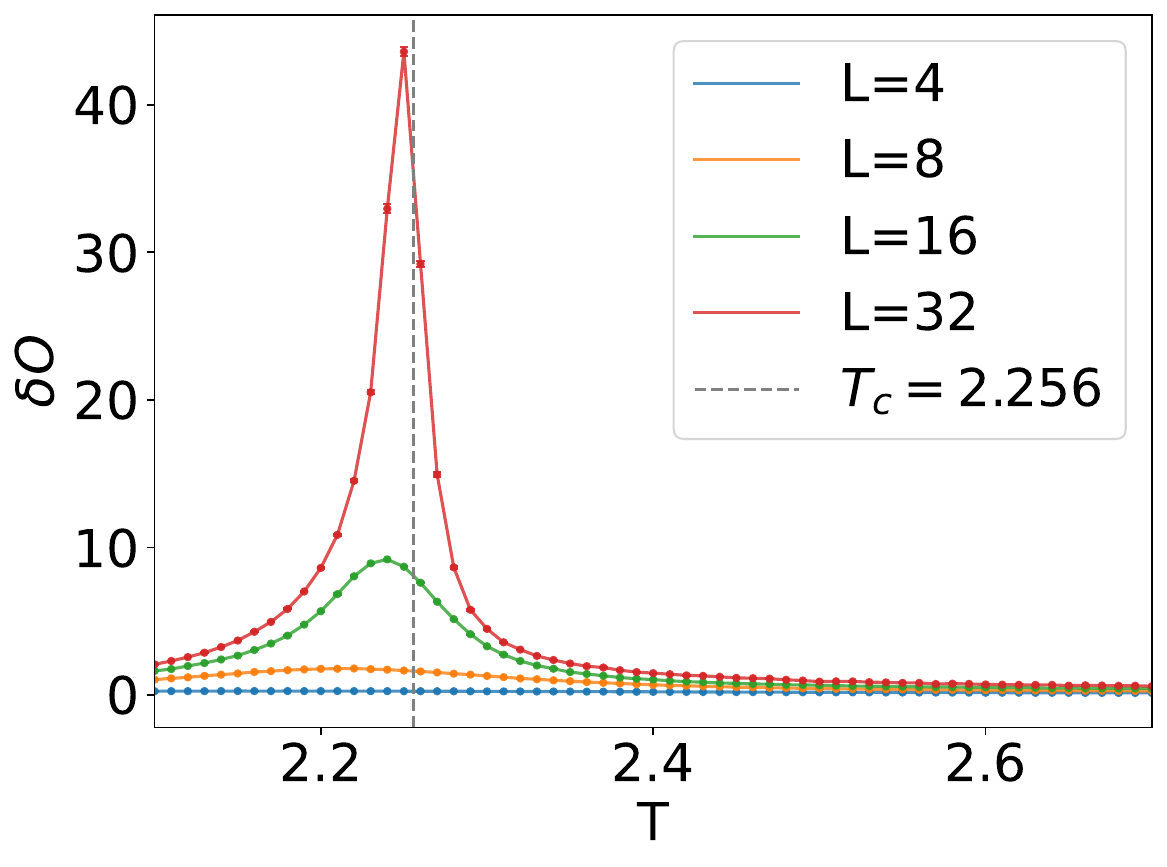} 
\par
(b)
\end{minipage}
\caption{The temperature dependence of the maximal row correlation $O$ (a), and its fluctuations $\delta O$ (b) for the $q=2$ FM Potts model on the 3D cubic lattice with various sizes. The vertical dashed lines in (a) and (b) mark the position of the critical temperature $T_c$ obtained in Ref.~\cite{Ferrenberg_2018}.}
\label{fig_square_3D_fm_Q2}
\end{figure}

\begin{figure}
\centering
\begin{minipage}{0.23\textwidth}
\centering
\includegraphics[width=\textwidth]{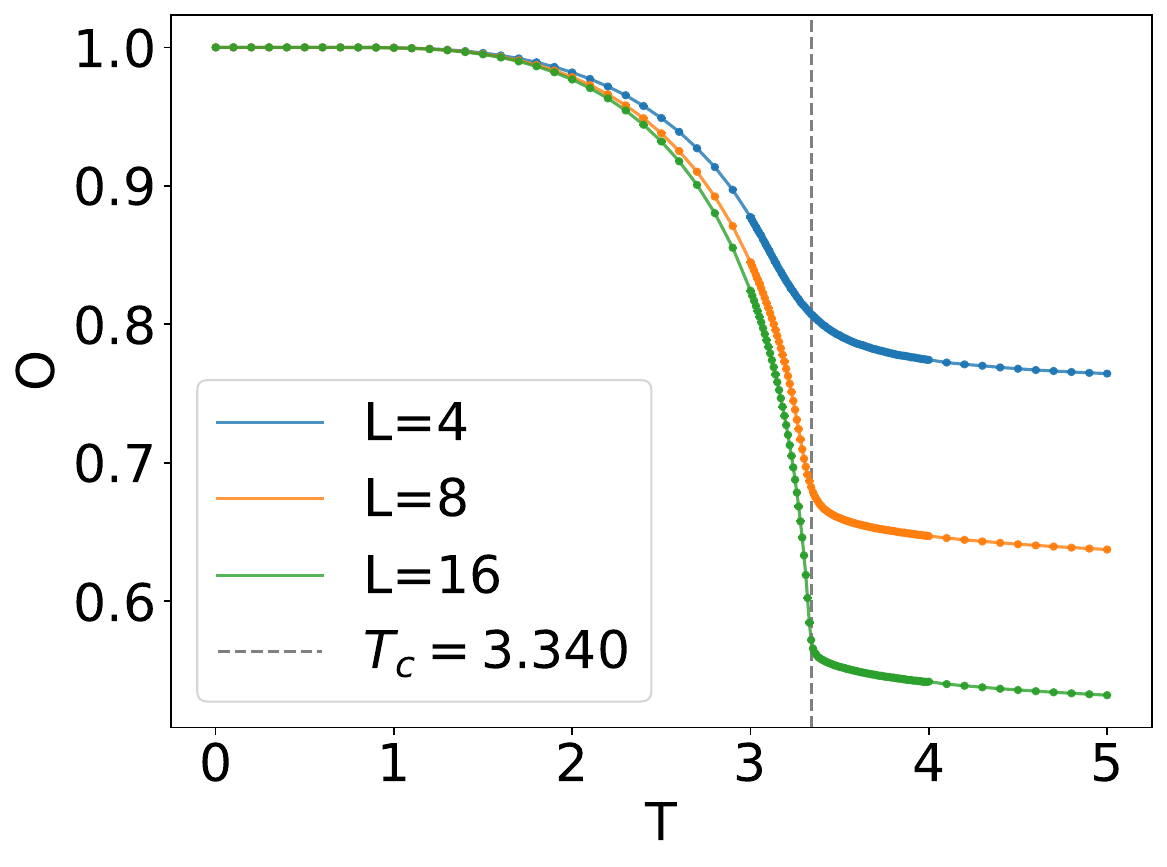} 
\par
(a)
\end{minipage}
\hfill
\begin{minipage}{0.23\textwidth}
\centering
\includegraphics[width=\textwidth]{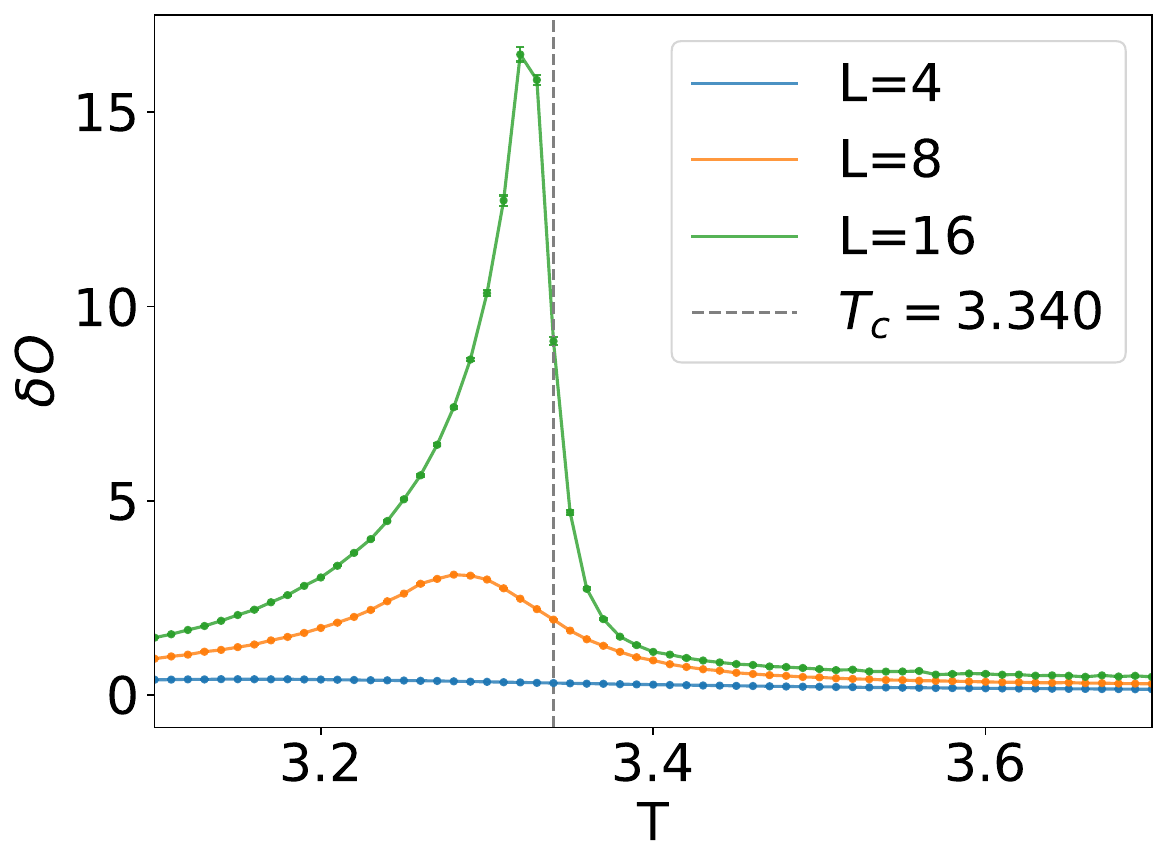} 
\par
(b)
\end{minipage}
\caption{Same quantities as in Fig.~\ref{fig_square_3D_fm_Q2}, but for the $q=2$ FM Potts model on the 4D cubic lattice. The vertical dashed lines in (a) and (b) mark the position of the critical temperature $T_c$ obtained in Ref.~\cite{Lundow_2009}.} 
\label{fig_square_4D_fm_Q2}
\end{figure}

In this appendix, we apply our method to higher-dimensional cases with $D>2$. For simplicity and illustrative purposes, we consider only the FM Potts model with $q=2$---equivalent to the FM Ising model---on simple cubic lattices in three and four dimensions.

Let us first consider the case of 3D cubic lattice, where the original configuration matrix $A$ becomes a rank-3 tensor of dimensions $L\times L\times L$. By reshaping the tensor and merging the last two indices into a single one, the tensor can be recast as a matrix of size as $L\times L^2$. The row correlation matrix $M$ is then constructed as before, via $M=A\,A^{\mathrm{T}}$, yielding a matrix of dimension $L\times L$. Strictly speaking, the matrix $M$ in this context on longer describes correlations between rows, but rather correlations between planes. Nevertheless, for notational consistency, we continue to refer to $M$ as the row correlation matrix, and the corresponding order parameter $O$ is likewise referred to as the maximal row correlation. 

It is important to note that the definition of $O$ requires a slight modification in three dimensions as $O=\sqrt[3]{\lambda_{\mathrm{M}}}/L$, i.e., the cubic root of the largest eigenvalue of $M$, divided by the matrix size $L$. This adjustment is motivated by the fact that, for a fully ordered configuration, the elements of $M$ scale as $L^2$, in contrast to the $L$ scaling in the 2D case (cf. Eq.~(\ref{eq_M_ordered})). An analogous procedure applies in 4D, where the last three indices are combined into one, and the order parameter is correspondingly defined as $\sqrt[4]{\lambda_{\mathrm{M}}}/L$.

Our MC results for the $q=2$ FM Potts model on 3D and 4D cubic lattices are shown in Figs.~\ref{fig_square_3D_fm_Q2} and \ref{fig_square_4D_fm_Q2}, where the temperature dependences of $O$ and its fluctuation $\delta O$ are plotted for system size up to $L=16$. We see that in both cases, $O$ converges uniformly to unity as $T\to 0$, indicating a fully ordered state at zero temperature. As $L$ increases, the peak in $\delta O$ sharpens and approaches the critical temperature $T_c$ previously determined by large-scale MC simulations~\cite{Ferrenberg_2018,Lundow_2009}. Given the limited system sizes accessible in this study---and because a precise determination of critical properties of these models is beyond our present scope---we did not attempt a data collapse to extract $T_c$ or the critical exponents $\nu$ and $\beta$. 

% (3D) Kc=0.221654628(5) Tc=2.25576161; (4D) Kc=0.1496947 Tc=3.3401316 

\section{Continuous spin models}\label{app_continuous}

\begin{figure}
\centering
\begin{minipage}{0.23\textwidth}
\centering
\includegraphics[width=\textwidth]{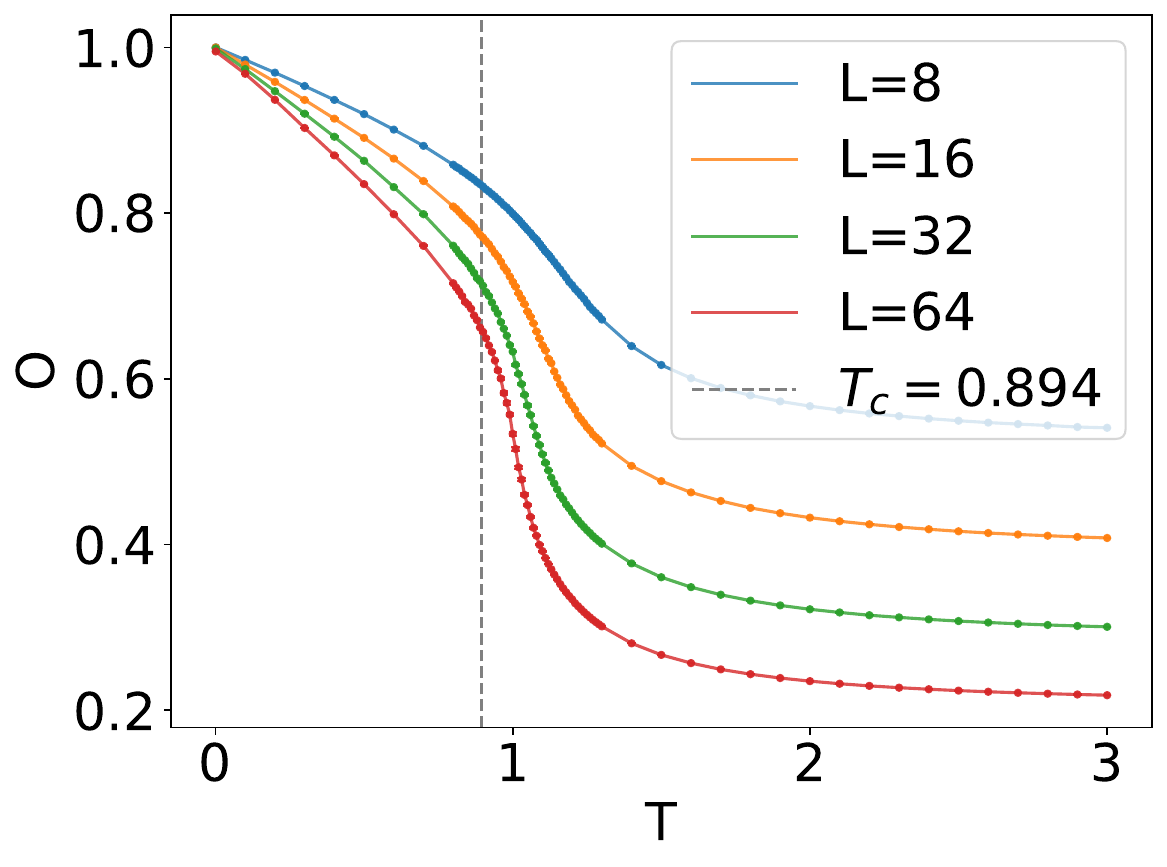} 
\par
(a)
\end{minipage}
\hfill
\begin{minipage}{0.23\textwidth}
\centering
\includegraphics[width=\textwidth]{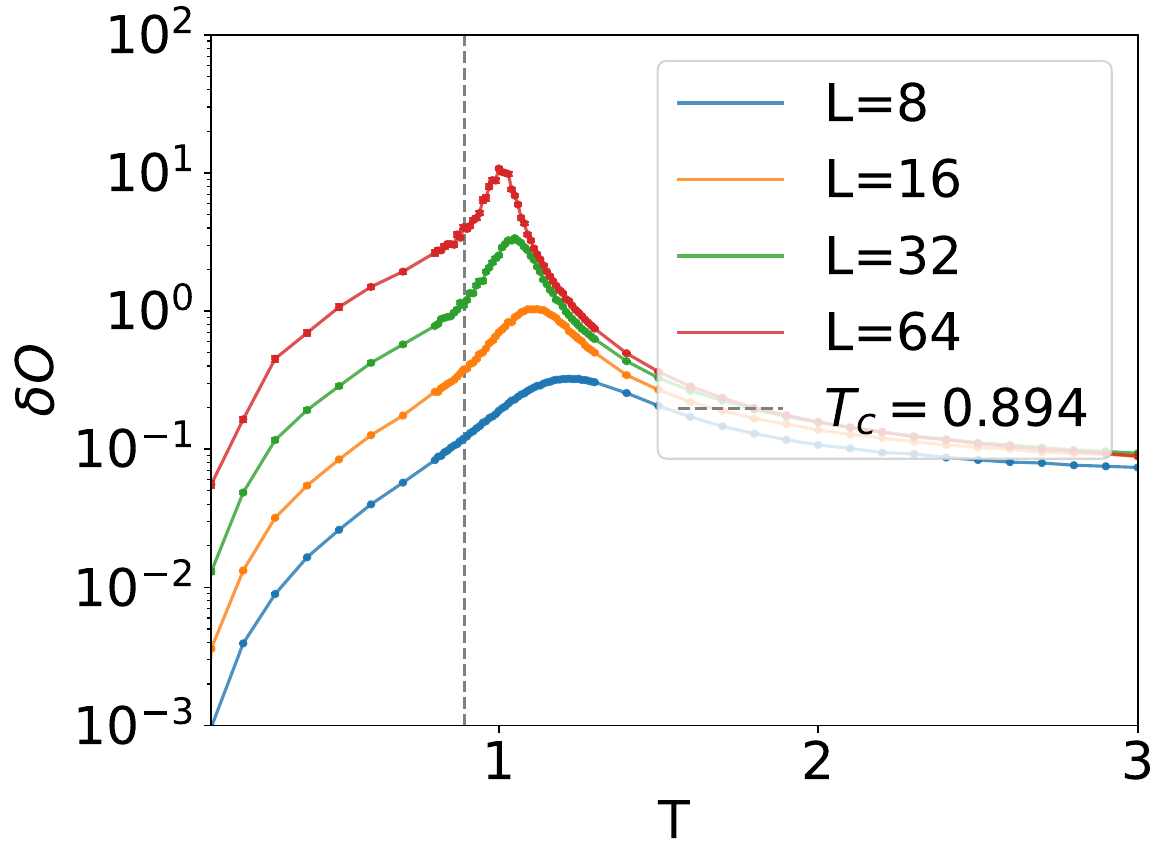} 
\par
(b)
\end{minipage}
\caption{Same quantities as in Fig.~\ref{fig_square_3D_fm_Q2}, but for the 2D XY model on the square lattice. Note that the fluctuation $\delta O$ shown in (b) is plotted on a logarithmic scale. The vertical dashed lines in (a) and (b) mark the position of the critical temperature $T_c$ obtained in Ref.~\cite{Gupta_1992}.}
\label{fig_square_fm_XY}
\end{figure}

\begin{figure}
\centering
\begin{minipage}{0.23\textwidth}
\centering
\includegraphics[width=\textwidth]{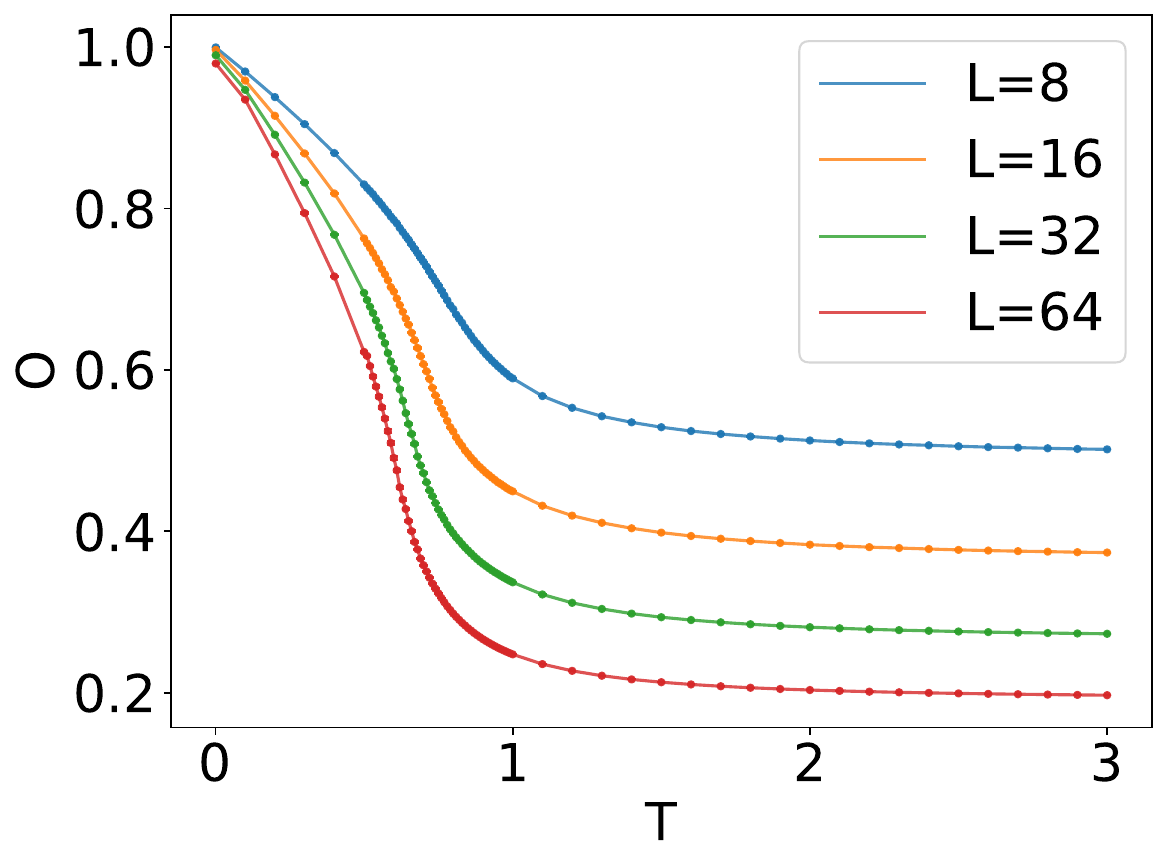} 
\par
(a)
\end{minipage}
\hfill
\begin{minipage}{0.23\textwidth}
\centering
\includegraphics[width=\textwidth]{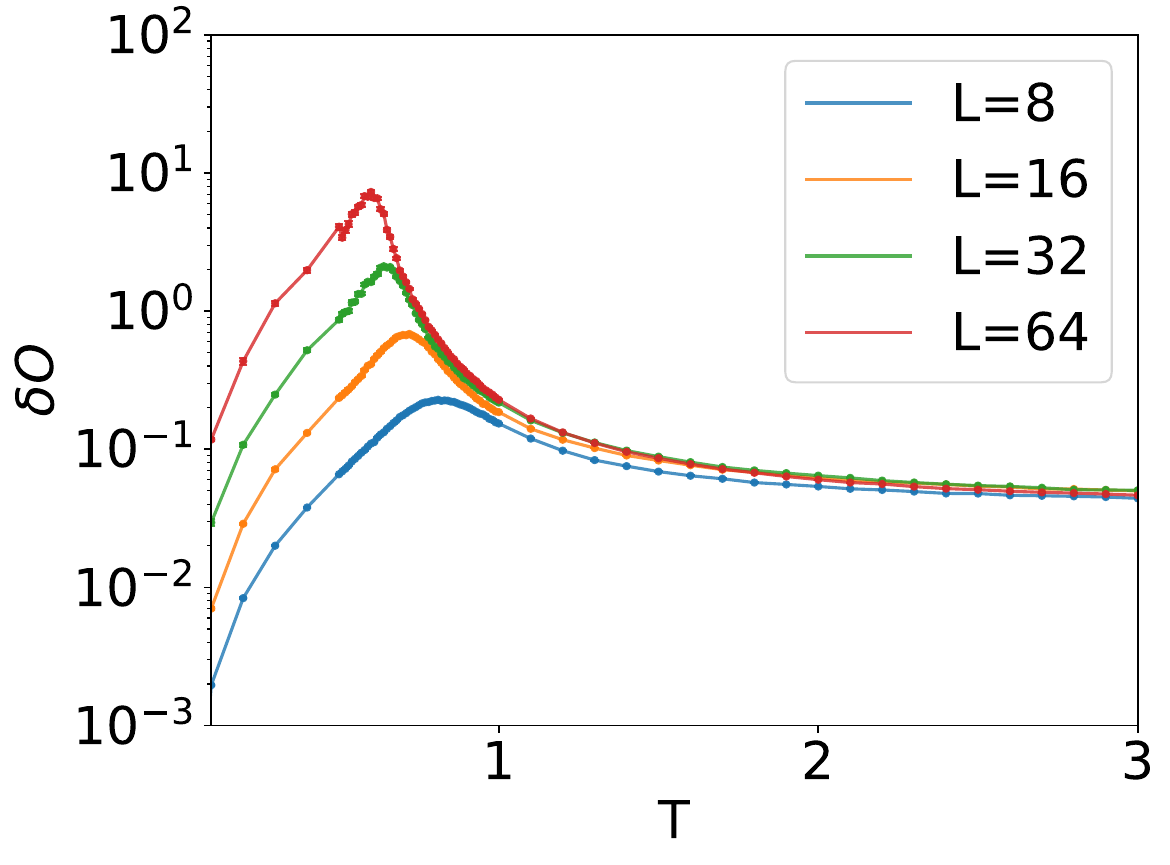} 
\par
(b)
\end{minipage}
\caption{Same quantities as in Fig.~\ref{fig_square_fm_XY}, but for the 2D classical Heisenberg model on the square lattice.} 
\label{fig_square_fm_Heisenberg}
\end{figure}

In this appendix, we study two paradigmatic models with continuous spin variables on the square lattice: the celebrated XY model and the classical Heisenberg model. The XY model undergoes a Berezinskii-Kosterlitz-Thouless (BKT) phase transition~\cite{Berezinskii_1971,*Kosterlitz_1973,*Kosterlitz_1974} at a critical temperature $T_c\approx 0.894$~\cite{Gupta_1992}, separating a low-temperature phase with quasi-long-range order from a high-temperature disordered phase characterized by an exponential decay of correlations. In contrast, for the 2D classical Heisenberg model, the existence of a BKT-like transition at finite temperature has been suggested but remains a subject of ongoing debate~\cite{Kapikranian_2007,Schmoll_2021,Oshikawa_2022}.

Figures~\ref{fig_square_fm_XY} and \ref{fig_square_fm_Heisenberg} present preliminary MC results for the XY and classical Heisenberg model on the square lattice, where we compute the temperature dependence of the order parameter $O$ and its fluctuations $\delta O$ for system size up to $L=64$. A notable feature in the behavior of $O$ is observed: unlike previous cases involving systems with fully or partial orders at low temperatures---where $O$ gradually saturates to a finite value and exhibits a clear plateau---here, although $O$ approaches to unity as $T\to 0$, it does so with a significant slope. We interpret this as an indication of the absence of long-range order at any nonzero temperature, consistent with the celebrated Mermin–Wagner–Hohenberg theorem~\cite{Mermin_1966,*Hohenberg_1967}. Moreover, the XY and Heisenberg models exhibit remarkable similar behavior. We note that our MC results for $O$ and $\delta O$ closely resemble those previously reported for the magnetization $M$ and susceptibility $\chi$~\cite{Kapikranian_2007}. It would be worthwhile to pursue more systematic studies at larger system sizes in the future to further substantiate these observations. 

% \bibliographystyle{apsrev4-2}
% \bibliography{lt.bib}


\end{document}